\documentclass[prl, twocolumn,superscriptaddress]{revtex4}
\usepackage{amsmath,amssymb,bm}
\usepackage{graphicx}
\usepackage{epstopdf}
\usepackage{latexsym}
\usepackage{subfigure}
\usepackage{color}
\usepackage{natbib}
\usepackage{braket}
\usepackage{hyperref}
\hypersetup{
  colorlinks,
  citecolor=magenta,
  linkcolor=blue,
  urlcolor=blue}
      
\DeclareMathAlphabet{\mathbbold}{U}{bbold}{m}{n}

\newcommand{\nccp}[1]{nc-$\mathbb{CP}^{#1}$}
\newcommand{\cp}[1]{$\mathbb{CP}^{#1}$}
\newcommand{\chiN}{\chi_{_\text{N}}}
\newcommand{\chiV}{\chi_{_\text{V}}}
\newcommand{\etaN}{\eta_{_\text{N}}}
\newcommand{\etaV}{\eta_{_\text{V}}}

\begin{document}

\title{Fate of $\mathbb{CP}^{N-1}$ fixed points with $q$-monopoles}

\author{Matthew S. Block}
\affiliation{Department of Physics \& Astronomy, University of Kentucky, Lexington, KY 40506-0055}

\author{Roger G. Melko}
\affiliation{Department of Physics and Astronomy, University of Waterloo, Ontario, N2L 3G1, Canada, \\and
	Perimeter Institute for Theoretical Physics, Waterloo, Ontario N2L 2Y5, Canada }

\author{Ribhu K. Kaul}
\affiliation{Department of Physics \& Astronomy, University of Kentucky, Lexington, KY 40506-0055}

\begin{abstract}
  We present an extensive quantum Monte Carlo study of the N\'eel-valence bond solid (VBS)
  phase transition on rectangular and
  honeycomb lattice SU($N$) antiferromagnets in sign problem free
  models. We find that in contrast to the honeycomb lattice and previously studied square lattice systems, on the rectangular lattice for small $N$ a
  first order N\'eel-VBS  transition is realized. On increasing $N\geq 4$, we observe that the
  transition becomes continuous and with the {\em same} universal
  exponents as found  on the honeycomb and square lattices (studied here for $N=5,7,10$), providing strong support for a deconfined quantum critical point. Combining
 our new results with previous numerical and analytical studies we
  present a general phase diagram of the stability of $\mathbb{CP}^{N-1}$ fixed points with $q$-monopoles. 
\end{abstract}
\date{\today}
\maketitle

The study of quantum critical points (QCP) has seen a lot of excitement in
both recent theoretical~\cite{xu2012:qcp} and experimental work~\cite{coldea2010:e8,zhang2012:qcp}. The most novel QCPs are those that do not have simple classical
analogues in one higher dimension. One of the most prominent examples of such a QCP is the direct continuous ``deconfined''  critical point (DCP) between N\'eel and valence-bond solid (VBS) phases in bipartite SU($N$) antiferromagnets~\cite{senthil2004:science}. Both states of matter are characterized by conventional broken symmetries, the N\'eel state by SU($N$) symmetry breaking and the VBS by lattice symmetry breaking.  A naive application of Landau theory would predict that since the two phases break distinct symmetries, a direct  N\'eel-VBS transition cannot be continuous. However by a subtle conspiracy of quantum interference and deconfinement, it has been shown that a continuous transition beyond the Landau paradigm can occur~\cite{senthil2004:deconf_long}. While the deconfined theory is by itself speculative (a ``scenario''), the discovery of sign-problem free models has allowed for unbiased tests by quantum Monte Carlo of the theoretical proposal on large two-dimensional lattice models, in a way unprecedented for an exotic quantum critical phenomenon~\cite{kaul2013:qmc}. 

\begin{figure}[t]
\centerline{\includegraphics[width=0.8\columnwidth]{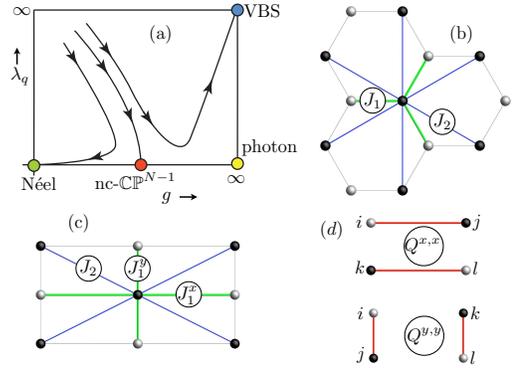}}
\caption{(color online). (a) Deconfined RG flow diagram for SU($N$) antiferromagnets with $q$-fold degenerate VBS phases, in the field theoretic space of monopole fugacity for $q$-monopoles ($\lambda_q$) and the tuning parameter $g$ of the critical point. (see text and \cite{senthil2005:jpsj} for details). In this work we give a complete phase diagram in $q$-$N$ space for which this RG flow diagram can be realized (see Table~\ref{tab:qN}).  (b,c,d) Couplings of Eq.~(\ref{eqn:fullmodel}): (b) The honeycomb lattice with $J_1$ and $J_2$. (c) The rectangular lattice with $J^x_1,J^y_1$ and $J_2$. (d) The $Q$ interaction shown is used here only on the rectangular systems. The A and B sublattices (black and white sites) have SU($N$) spins transforming in the fundamental and conjugate to fundamental representations, respectively. }
\label{fig:lattices}
\end{figure}
The speculative assumptions that underlie the DCP concept concern the existence and stability of certain critical fixed points. The DCP idea builds on 
the $\mathbb{CP}^{N-1}$ description of bipartite two-dimensional SU($N$) quantum antiferromagnets~\cite{read1990:vbs}. The \cp{N-1} field theory consists of $N$ complex scalars $z_\alpha$ interacting with a U(1) gauge field $a_\mu$.
Destructive
interference from Berry phases result in the suppression of
monopoles in $a_\mu$ unless they have a charge, $q$~\cite{haldane1988:berry}. A central result is that $q$
in the simplest cases (of interest here) is equal to the degeneracy of the VBS phase~\cite{read1990:vbs}, so the square lattice has $q=4$, the honeycomb $q=3$ and the rectangular lattice has $q=2$. 
 The discussion so far is on firm grounds.
The two speculative ingredients that allow for a deconfined quantum critical point between N\'eel and VBS
states in SU($N$) antiferromagnets on lattices with $q$-fold
degenerate VBS state are: (1) the existence of a critical fixed
point in the ``non-compact'' monopole-free \cp{N-1} theory~\cite{motrunich2004:hhog} (this will be referred to as \nccp{N-1}), and, (2) the ``dangerous irrelevance'' of
$q$-monopole insertions at the \nccp{N-1} fixed point. If these two conditions are met, the resulting
``deconfined'' renormalization group flow diagram~\cite{senthil2005:jpsj} is as shown in Fig.~\ref{fig:lattices}
(a).

\begin{figure}[t]
\centerline{\includegraphics[width=\columnwidth]{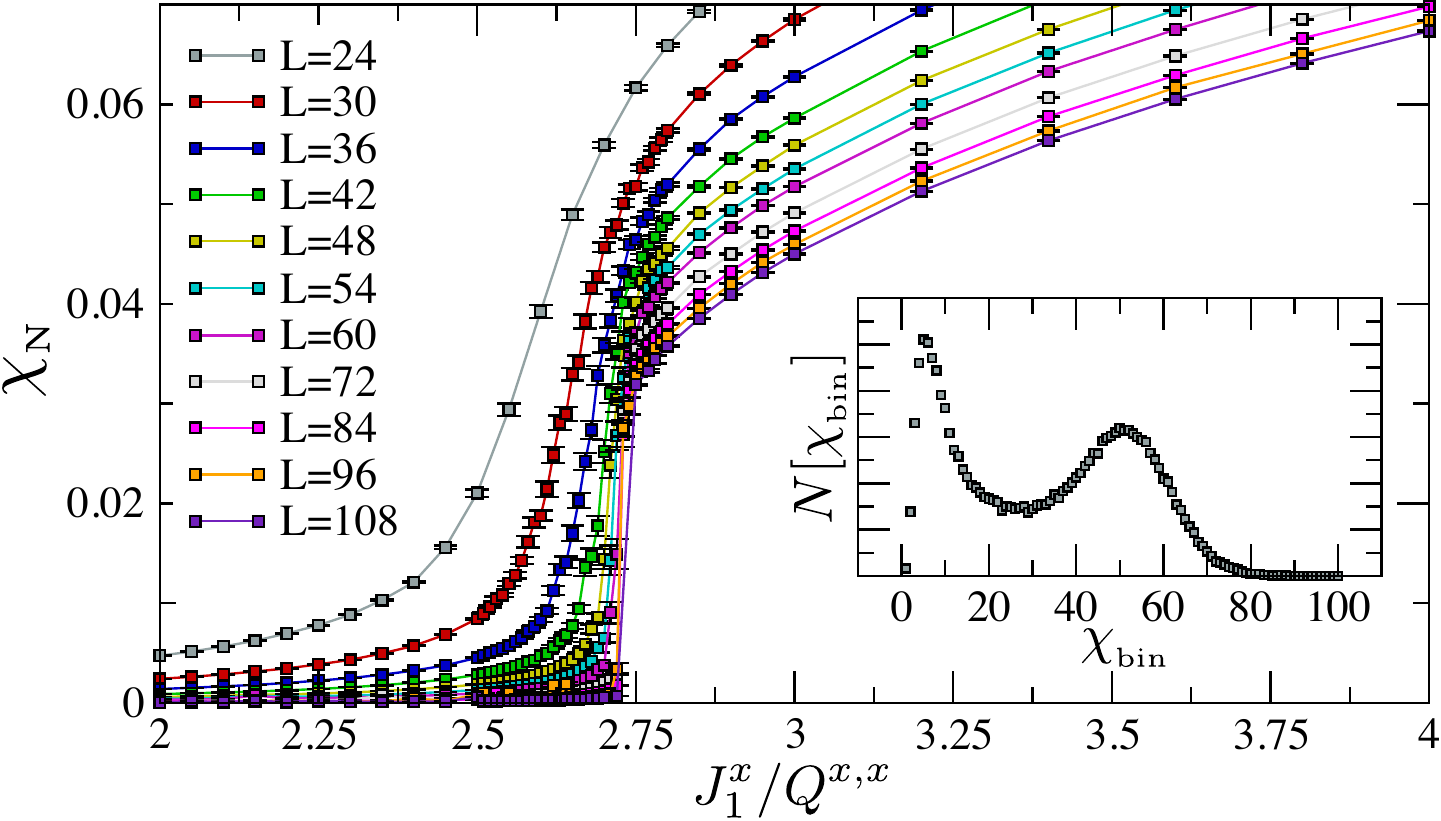}}
\caption{(color online).  First order transition for $q=2$ and $N=3$ (rectangular lattice with SU(3) spins). Magnetic susceptibility for SU(3) on the rectangular lattice.  The sharp jump is indicative of a first-order transition.  The inset shows a double peaked histogram of data taken from a point in the middle of the transition ($J_1^x/Q^{x,x}=2.71$) for $L=48$ thus providing further evidence for a first order transition.  To accommodate the rectangular lattice symmetry~\cite{sandvik1999:aspectratio}, we take a lattice with $L_x=4L_y/3$; $L=L_y$ in our legend.}
\label{fig:1stordersu3}
\end{figure}

The most extensive studies of deconfined criticality in microscopic models have focussed on the case $N=2$ and $q=4$~\cite{sandvik2007:deconf, melko2008:jq, jiang2008:first,sandvik2010:log, banerjee2010:log} ({\em i.e.} the square lattice with SU(2) spins). Other studies have tackled the cases $q=4,2\leq N\leq 12$~\cite{lou2009:sun,kaul2011:su34,kaul2012:j1j2,banerjee2010:su3} (square lattice with SU($N$) spins) and $N=2, q=3$~\cite{damle2013:jqhcsu2} (the honeycomb lattice with SU(2) spins). The nature of the transition in the $q=\infty$ limit for $N=2$ by studying the classical \nccp{N-1} model in three dimensions has been debated extensively~\cite{motrunich2008:cp1,kuklov2008:first,noguiera2007:dcp,chen2013:first}. We shall extend the studies of deconfined criticality by studying the case $q=2$ (rectangular lattice) and $q=3$ (honeycomb) for $N\leq 10$. Our main conclusions are as follows: We find clear evidence that the N\'eel-VBS transition on the rectangular lattice ($q=2$) is first order for $N=2,3$ and continuous for $N \geq 4$. We find the anomalous dimensions ($\etaN$ and $\etaV$) for $N=5,7,10$ are in agreement with each other on the rectangular ($q=2$), honeycomb lattices ($q=3$) and square lattices ($q=4$), all of which are consistent with the analytic $1/N$ expansion for the \nccp{N-1} model ($q=\infty$) (see  Fig.~\ref{fig:etacomp}). Finally, combining our new results with existing work, we suggest a general phase diagram for the values of $N$ and $q$ for which the deconfined RG flow in Fig.~\ref{fig:lattices}(a) is realized and a continuous deconfined N\'eel-VBS transition can occur (see Table~\ref{tab:qN}). 
 
\begin{figure}[t]
\centerline{\includegraphics[width=\columnwidth]{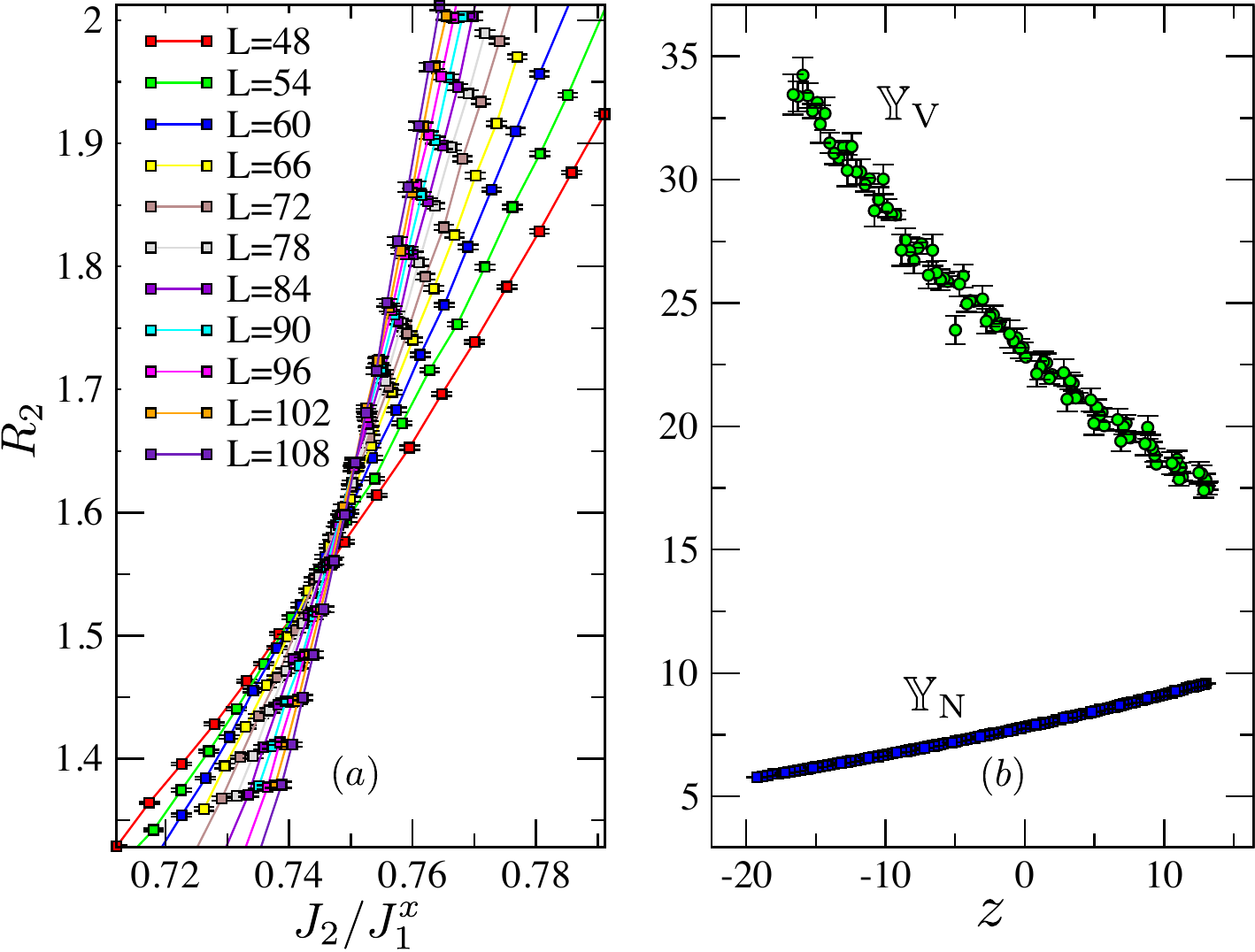}}
\caption{(color online).  Continuous transition for $q=2$ and $N=7$ (rectangular lattice with SU(7) spins). (a) This panel shows the Binder ratio data. (b) Both the magnetic (blue squares) and VBS (green circles) susceptibility data.  The data has been collapsed such that $\mathbb{Y}_{\text{N}}(z)=L^{1+\etaN}\chiN(z)+(a+bz)L^{-\omega}$ and $\mathbb{Y}_{\text{V}}(z)=L^{1+\etaV}\chiV(z)$ with $\etaN=0.639$, $a=8.5$, $b=0.1$, $\omega=0.5$, and $\etaV=1.26$.  Also, $z=[(g-g_c)/g_c]L^{1/\nu}$ with $g=J_2/J_1^x$, $g_c=0.7552$ and $\nu=0.69$.  For the magnetic susceptibility, the following system sizes were used in the collapse: $L=42,48,54,60,66,72,78,84,90,96,102,108$.  For the VBS susceptibility, the following system sizes were used in the collapse: $L=36,42,48,54,60,66$.}
\label{fig:chianrecsu7}
\end{figure}

{\em Model:}
We consider bipartite SU($N$) antiferromagnets in which the spins on the A sublattice transform under the fundamental representation of SU($N$) while those on the B sublattice transform under the conjugate to the fundamental representation used fruitfully in both past analytic~\cite{affleck1985:lgN,read1989:vbs} and numerical~\cite{harada2003:sun,beach2009:sun} studies.  Following previous work reviewed in detail in Ref.~\cite{kaul2013:qmc}, we can construct sign-problem free Hamiltonians that maintain the SU($N$) symmetry from two operators, a projection operator: $\mathcal{P}_{ij}=\sum_{\alpha,\beta=1}^N\Ket{\alpha\alpha}_{ij}\Bra{\beta\beta}_{ij}$ (with $i$ and $j$ on opposite sublattices) and a permutation operator: $\Pi_{ij}=\sum_{\alpha,\beta=1}^N\Ket{\alpha\beta}_{ij}\Bra{\beta\alpha}_{ij}$ (with $i$ and $j$ on the same sublattice). The Hamiltonian we will study can be written in the following very general form,
\begin{equation}
\label{eqn:fullmodel}
H=-\sum_{i,j}\frac{J^{ij}_1}{N}\mathcal{P}_{ij}-\sum_{i,j}\frac{J^{ij}_2}{N}\Pi_{ij}-\sum_{\text{pl}}\frac{Q^{ij,kl}}{N^2}\mathcal{P}_{ij}\mathcal{P}_{kl}.
\end{equation}
Illustrations of how each of the terms appears is shown in Fig.~\ref{fig:lattices} (b,c,d). For small $N$ the $J_1$ only models are always N\'eel ordered and for large-$N$ they are always VBS ordered.  To study the N\'eel-VBS transition at fixed $N$, we use the $J_2$ and $Q$ terms. As studied previously, the $J_2$ interaction strengthens the N\'eel state by favoring ferromagnetic order on each of the sublattices~\cite{kaul2012:j1j2}, while the $Q$ interaction favors the VBS phase by preferring the plaquettes to enter singlet states~\cite{sandvik2007:deconf}. With the Hamiltonian so defined we can study all the N\'eel-VBS phase transitions of interest, as we detail below. We shall study the model Hamiltonian using the unbiased and powerful stochastic series expansion quantum Monte Carlo method~\cite{sandvik2010:vietri}. Details of the observables are provided in the Supplementary Materials (SM).

\begin{figure}[t]
\centerline{\includegraphics[width=\columnwidth]{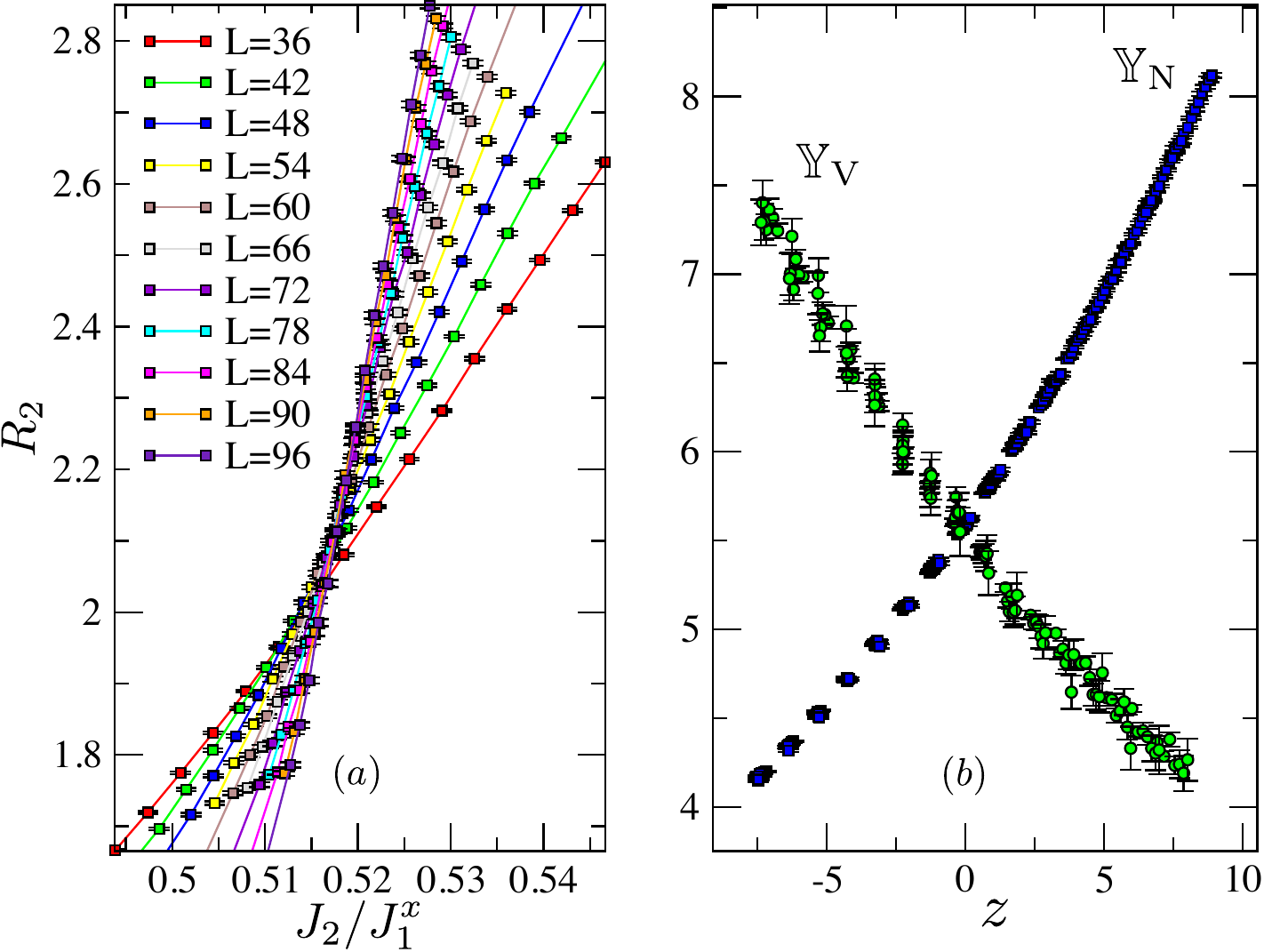}}
\caption{(color online).  Continuous transition for $q=3$ and $N=7$ (honeycomb lattice with SU(7) spins). (a) The Binder ratio. (b) Both the magnetic (blue squares) and VBS (green circles) susceptibility data.  The data has been collapsed such that $\mathbb{Y}_{\text{N}}(z)=L^{1+\etaN}\chiN(z)+(a+bz)L^{-\omega}$ and $\mathbb{Y}_{\text{V}}(z)=L^{1+\etaV}\chiV(z)$ with $\etaN=0.67$, $a=20.0$, $b=0.8$, $\omega=1.0$, and $\etaV=1.41$.  Also, $z=[(g-g_c)/g_c]L^{1/\nu}$ with $g=J_2/J_1$, $g_c=0.5196$ and $\nu=0.72$.  For the magnetic susceptibility, the following system sizes were used in the collapse: $L=36,42,48,54,60,66,72,78,84,90,96$.  For the VBS susceptibility, the following system sizes were used in the collapse: $L=18,24,30,36,42,48,54$.  There are $2L^2$ lattice sites.}
\label{fig:chihcsu7}
\end{figure}

{\em Rectangular Lattice:}
We begin by studying the phase transition between the N\'eel state and a $q=2$-fold degenerate VBS as a function of $N$.  We study Eq.~(\ref{eqn:fullmodel}) on a rectangular lattice (see Fig~\ref{fig:lattices}(b)), where the couplings are chosen to have rectangular symmetry, {\em i.e.} invariant under translation in $x$ and $y$, but break the $\pi/2$ rotation symmetry that would be present on a square lattice. On such a lattice the VBS state must be two fold degenerate, achieving $q=2$~\cite{harada2007:deconf}. Specifically, we begin by taking $J^y_{1}=0.8J^x_{1}$. For these couplings the model is N\'eel order for $N\leq 4$ and VBS ordered for $N>4$ (see SM for details). To study the N\'eel-VBS transition for $N\leq 4$ we add a $Q$ interaction (here we use $Q^{y,y}=0.8Q^{x,x}$) and tune the ratio $J^x_1/Q^{x,x}$.  Remarkably, we find first-order transitions for $N=2,3$ (see Fig.~\ref{fig:1stordersu3}) and a continuous transition for $N=4$ (see SM).   For $N>4$ we can study the N\'eel-VBS transition by introducing a $J_2$ coupling. For all $N>4$ we find strong evidence for a continuous transition. A sample of our data for $N=7$ is shown in Fig.~\ref{fig:chianrecsu7} (additional data for $N=5,10$ are shown in SM). Although we note that in principle our finding of a first order transition cannot rule out a continuous transition in another model with the same $q,N$,  it is natural to assume that the first order transition observed for $q=2$ is generic and results from the relevance of $\lambda_2$ for $N=2,3$.  This assumptions lends itself naturally to an interesting interpretation of our numerical observation that for $q=2$ the transition is first order for $N=2,3$ and continuous for $N\geq 4$: in general we expect that for a fixed $q$ the scaling dimension of the monopole operator should increase as $N$ increases~\cite{murthy1990:mono}. What we have observed here then is that for $q=2$ the scaling dimension is large enough to become irrelevant only when $N\geq 4$ [in agreement with the RG flow in Fig.~\ref{fig:lattices}(a)], but for $N=2,3$ the operator is a relevant perturbation [in contradiction to the RG flow shown in Fig.~\ref{fig:lattices}(a)] and thus drives the transition first order. 
\begin{figure}[t]
\centerline{\includegraphics[width=\columnwidth]{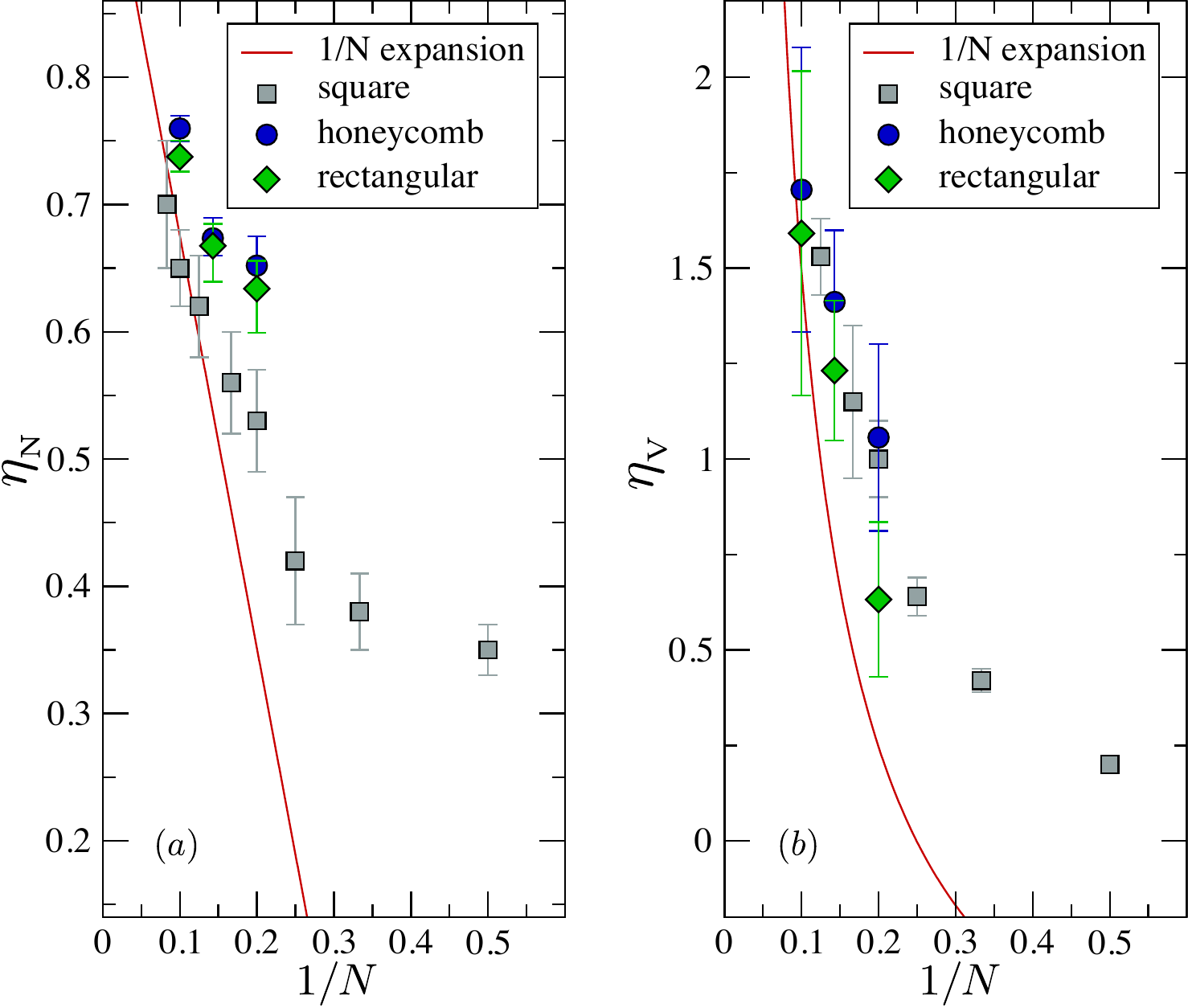}}
\caption{(color online).  Comparison of anomalous dimensions of N\'eel and VBS operators in the case of continuous transitions for $q=2,3$ and $4$. (a) Anomalous dimension of the N\'{e}el order parameter as a function of $1/N$.  (b) Anomalous dimension of the VBS order parameter as a function of $1/N$.  The gray squares are the results of a previous square lattice study ($q=4$) ~\cite{lou2009:sun,kaul2012:j1j2}. The blue circles are new results from the honeycomb lattice ($q=3$) and the green diamonds are new results from the rectangular lattice ($q=2$).  The red line is the $1/N$ expansion. The agreement of the new data with both the $q=4$ data as well as the $1/N$ computation is striking.}
\label{fig:etacomp}
\end{figure}

{\em Honeycomb lattice:} Next, we study the case of a $q=3$-fold degenerate valence bond solid phase.  We can achieve this by studying our model, Eq.~(\ref{eqn:fullmodel}), on the honeycomb lattice [see Fig.~\ref{fig:lattices}(a)].  The case of SU(2), SU(3) and SU(4) have recently been studied~\cite{damle2013:jqhcsu2,harada2013:deconf} and the transition was shown to be continuous and is expected to remain continuous for larger $N$~\cite{murthy1990:mono}. Our goal is to verify this expectation by studying the QCP for large-$N$ and extract $\etaN$ and $\etaV$ at the critical point for $N=5,7,10$.  Our starting point now is a $J_1$ only model on the nearest neighbors of a honeycomb lattice, which is VBS ordered for $N=5,7,10$ (see SM for a full study of the $J_1$ model as a function of $N$). To tune into the N\'eel state we introduce a $J_2$ between second nearest neighbors on the honeycomb. We observe very good evidence for a continuous transition; a sample of our data for $N=7$ is shown in  Fig.~\ref{fig:chihcsu7}.

\begin{table}[t]
\begin{tabular}{||c||c|c|c|c|c|c|c||} 
\hline 
\hline
 $N=\infty, 1/N$ & $I$ & $I$ & $I$ & $I$ & $\dots$& $I$   &  nc-$\mathbb{CP}^{N-1}$\\
\hline
\hline
$\dots$ & $$ & $$ & $$ & $$ & $$& $$  & $$\\
\hline
 $N=10$ & $R$ & $I$ & $I$ & $I$ & $$  & $I$&  nc-$\mathbb{CP}^{9}$\\
\hline
 $N=9$ & $R$ & $I$ & $I$ & $I$ & $$  & $I$&  nc-$\mathbb{CP}^{8}$\\
\hline
 $N=8$ & $R$ & $I$ & $I$ & $I$ & $$  & $I$&  nc-$\mathbb{CP}^{7}$\\
\hline
 $N=7$ & $R$ & $I$ & $I$ & $I$ & $$  & $I$&  nc-$\mathbb{CP}^{6}$\\
\hline
 $N=6$ & $R$ & $I$ & $I$ & $I$ & $$  & $I$&  nc-$\mathbb{CP}^{5}$\\
\hline
 $N=5$ & $R$ & $I$ & $I$ & $I$ & $$  & $I$&  nc-$\mathbb{CP}^{4}$\\
 \hline
 $N=4$ & $R$ & $I$ & $I$ & $I$ & $$  & $I$&  nc-$\mathbb{CP}^{3}$\\
 \hline
 $N=3$ & $R$ & $R$ & $I$ & $I$ & $$ & $I$&  nc-$\mathbb{CP}^{2}$ \\
 \hline
$N=2$ & $R$ & $R$ & $I$ & $I$ & $$ & $I$& nc-$\mathbb{CP}^{1}$\\
\hline 
$N=1$ & $R$ & $R$ & $R$& $I$ & $$& $I$ & $XY$\\
\hline
$N=0$ & $R$ & $R$ & $R$& $R$ & $$& $R$ & photon\\
\hline
$$ & $q=1$ & $q=2$ & $q=3$& $q=4$ & $\dots$ & $q=\infty$ & $$\\
\hline
\hline 
\end{tabular}
\caption{Table showing the inferred relevance ($R$) or irrelevance ($I$) of
  $q$-monopoles at the \nccp{N-1} fixed point, which our current study has allowed to complete. Numerical simulations of the N\'eel-VBS transition in the models discussed here only allow studies for $N\geq 2$.  The entries with $R$
  correspond to an unstable fixed point, and $I$ to a stable fixed
  point that can then support the RG flow of Fig.~\ref{fig:lattices}(a). At some currently unknown critical value of $N>10$, the $q=1$
  case switches from $R$ to $I$.  }
\label{tab:qN}
\end{table}

{\em Discussion:} In addition to the results already presented for SU(7), we have extracted $\etaN$, and $\etaV$, for $q=2,3$ and $N=5,10$.  Fig.~\ref{fig:etacomp} shows all of our results in comparison to previous data from the square lattice study~\cite{kaul2012:j1j2} and the analytic predictions~\cite{murthy1990:mono, metlitski2008:mono,kaul2008:u1}.  Our procedure for extracting the critical exponents, as well as the values of the critical couplings, is detailed in the SM. We find that within the error bars of our calculation, the anomalous dimension of the N\'eel and VBS order parameters are the same for rectangular, honeycomb and square lattice, which is strong evidence for the fact that the phase transition in these three different cases is controlled by the same fixed point. This must mean that the the lattice anisotropy is irrelevant for $N=5,7,10$, which in the field theory language corresponds to the irrelevance of 2,3 and 4-fold monopoles at these fixed points~\cite{senthil2005:jpsj}.  In addition we find that as $N$ increases the critical indices approach the value computed in the $1/N$ expansion in the \nccp{N-1} field theory, as shown in Fig.~\ref{fig:etacomp}. This is evidence that the common critical point is indeed the \nccp{N-1} theory as predicted by ``deconfined criticality.''

We now put our results in a broader context (see Table~\ref{tab:qN} and for a more detailed discussion, the SM). Since the critical theory of the SU($N$) N\'eel to $q$-fold degenerate VBS transition is described by the \cp{N-1} theory with $q$-monopoles, we can think of our numerical simulations of antiferromagnets as a way to learn about the \cp{N-1} theory with $q$-monopoles. The \nccp{N-1} fixed point is known to exist analytically at large-$N$~\cite{halperin1974:largeN} and for $N=1$~\cite{dasgupta1981:dual} (for $N=0$ there are no matter field and one has a stable photon phase). We shall take the point of view that by continuity it exists for all $N$, this is the right-most column of Table~\ref{tab:qN} (we note here that the case $N=2$ has been debated in the literature~\cite{motrunich2008:cp1,kuklov2008:first,noguiera2007:dcp,chen2013:first}). We can now ask whether $q$-monopoles are relevant (R) or irrelevant (I) at the \nccp{N-1} fixed point. Past analytic and field theoretic work have addressed the question for $N=0$~\cite{polyakov1987:gauge}$, N=1$~\cite{dasgupta1981:dual} and $N=\infty$~\cite{murthy1990:mono}. The column $q=1$ has recently been addressed in simulations of loop models~\cite{nahum2011:loops} and bilayer SU($N$) antiferromagnets~\cite{kaul2012:bilayer}. The column $q=4$ has been addressed by studying the critical point of the square-lattice N\'eel-VBS transition~\cite{kaul2012:j1j2}. Here we have provided the final piece of the puzzle by studying the $q=2$ and $q=3$ case (see~\cite{damle2013:jqhcsu2} for a study of $q=3,N=2$), where we have explicitly seen the change from a first order to a continuous transition as $N$ is increased for $q=2$. The rest of the table can be filled out by making the reasonable assumption that once an entry is $I$ it will stay $I$ for increasing $q$ or $N$. It is expected that the $q=1$ column will switch from $R$ to $I$ at some large finite value of $N$; this value has not been accessed in numerical simulations currently.

We gratefully acknowledge helpful discussions with M. Fisher and A. Sandvik. The research reported here was supported in part by NSF DMR-1056536 (MSB, RKK) and the Natural Sciences and Engineering Research Council of Canada (RGM)

\bibliography{/Users/rkk/Physics/PAPERS/BIB/career}

\newpage

\section{Supplementary Materials}

In this Supplementary Material, we present additional details of the measurements and analysis tools used to examine the properties of the phase transitions in the models described in the main article.

\section{Detailed Discussion of Table I}\label{sec:tab1}

The $\mathbb{CP}^{N-1}$ field theory in 2+1 dimension, describing $N$
complex bosonic fields, $z_\alpha$,
interacting with a U(1) gauge field, $a_\mu$, can be represented by the
following action,
\begin{equation}
\label{eq:cpN}
S = \frac{1}{g}\int d^3x \left ( \sum_{\alpha=1}^{N}| (\partial_\mu -i
  a_\mu)z_\alpha |^2  + F_{\mu\nu}F^{\mu\nu}\right ).
\end{equation}
with the constraint, $\sum_{\alpha=1}^N |z_\alpha|^2=1$. This field theory has a long and rich history
in condensed matter physics. It has been applied to a
wide variety of phase transitions including those in superconductors~\cite{halperin1974:largeN},
liquid crystals~\cite{degennes1993:lc}, loop models~\cite{nahum2011:loops}, and quantum
antiferromagnets~\cite{read1990:vbs,senthil2004:science}. It is also
amongst the simplest field theories that displays the
``Higgs'' phenomena~\cite{zee2010:nuts}.

Topological defects play a key role in the nature of phase
transitions~\cite{chaikin2000:cmp}. In a U(1) gauge theory in 2+1 dimensions the topological
defects are ``monopoles'' characterized by an integer $q$ which counts the number of units of flux
emanating from the point-like defect (we shall refer to these as $q$-monopoles). In this work we are interested in the role of these $q$-monopoles at the \cp{N-1} fixed point.

{\em Limiting cases:} We now turn to many limiting cases of our Table I which were known previously.

Before we consider what happens with $q$-monopoles, let us begin by considering the field theory, Eq.~(\ref{eq:cpN}),
without monopoles (i.e. by setting the monopole fugacity $\lambda_q=0$ for all $q$ in Fig. 1(a) of the main text). This can be achieved technically by making the gauge field
non-compact ({\em i.e.} $-\infty \leq a_\mu \leq \infty$), we shall call this the \nccp{N-1}
model~\cite{motrunich2004:hhog}. The model has two phases: a ``Higgs
phase'' where $z$ is condensed and the gauge field is hence massive, and a ``photon
phase'' where the $z$ field is massive and the gauge field
fluctuations are gapless; these states must be separated by a phase transition.     
At large-$N$ it has been shown
that the transition is continuous and its universal
properties can be computed in a $1/N$ expansion from the $N=\infty$
limit~\cite{halperin1974:largeN}. At $N=1$, a duality transformation has
shown that the  nc-$\mathbb{CP}^{0}$  model has a continuous
transition in the universality class of the
$XY$-model~\cite{dasgupta1981:dual}. It has been plausibly
hypothesized that the model continues to supports a second-order
transition between the limiting cases, {\em i.e.,} for all
$N$ between 1 and $\infty$~\cite{senthil2004:science}. 
Direct numerical simulations at $N=2$ have found good evidence for a
continuous transition~\cite{motrunich2008:cp1} (see
however~\cite{kuklov2008:first}). The case $N=0$ is just a pure
non-compact gauge theory that has a gapless photon phase. 
The nature of
the fixed points in the monopole-free ``non-compact'' theories are shown in the right most column of
Table~I.

Now imagine allowing $q$-monopole events at the \nccp{N-1} fixed point. If $q$
is made sufficiently large for any $N\neq 0$, it will clearly not affect the stability of the
monopole-free critical point and they are hence irrelevant ($I$ in
Table~I). This is
shown in the $q=\infty$ column of Table~I. The only exception is $N=0$ where the
introduction of monopoles always confines the photon phase~\cite{polyakov1987:gauge}.

Polyakov's confinement argument implies that the photon phase with no matter fields is always unstable to the
introduction of any $q$-monopoles. This is represented in the $N=0$ row of
Table~I.

Each of the entries in the $N=1$ row of Table~I can be
filled in using the power of the duality method. 
In the dual picture~\cite{dasgupta1981:dual} the critical point of the
\nccp{0} theory becomes an inverted $XY$ phase transition and the
$q$-monopoles become a $C_q$ magnetic field applied to the $XY$ order
parameter. It is well established that for $q\leq 3$ the $C_q$
perturbation is a relevant perturbation ($R$) at the $XY$ fixed point~\cite{fukugita1989:3pd3}
and for $q\geq 4$ the $C_q$~\cite{carmona2000:cubicanis}
perturbation is (dangerously) irrelevant ($I$) at the $XY$ fixed point. 

Finally the stability of the \nccp{N-1} fixed point to $q$-monopoles
has been studied in the large-$N$ limit~\cite{murthy1990:mono}, where
it has been shown that the monopole scaling dimension is proportional
to $N$. This renders monopoles irrelevant independent of $q$ at
large-$N$, as shown in the $N=\infty$  row of  Table~I. 

Beyond the cases discussed above, for finite-$q$ and finite-$N$, one must resort to numerical
simulations. Directly simulating the gauge theory Eq.~(\ref{eq:cpN}) with
constraints on the topological defects is notoriously
difficult. Instead an efficient approach we shall use here is to
study sign-problem free models of quantum antiferromagnets~\cite{kaul2013:qmc} and
exploit their close connection to the \cp{N-1} model with
$q$-monopoles~\cite{haldane1988:berry, read1990:vbs}.

The case of $q=4$ has been studied extensively by numerical
simulations~\cite{sandvik2007:deconf, melko2008:jq, lou2009:sun, kaul2012:j1j2}. In the language
of the quantum antiferromagnet this corresponds to the SU($N$) N\'eel-valence
bond solid transition on the square lattice~\cite{senthil2004:science}. From the numerical
studies there is strong evidence
here that the \nccp{N-1} fixed point is stable for all $N\geq 2$ at
$q=4$.

The case $q=1$ has been studied using the bilayer quantum
antiferromagnets and in loop models~\cite{kaul2012:bilayer, nahum2011:loops}, and there is clear evidence that for all $N\leq 10$
studied, monopole insertion is a relevant perturbation. Since in the
large-$N$ limit the single monopole operator becomes irrelevant there
must be some large finite value (currently unknown) at which the $q=1$
column switches from $R$ to $I$.

In order to complete the table we need to address what transpires at
$q=2$ and $q=3$ for each $N$. This has been described in detail in the text of the paper.

\section{Magnetic Quantities of Interest}\label{sec:magquant}
\subsection{Magnetic Susceptibility}\label{sec:magsus}
We begin by defining an SU($N$) generalization of the magnetic order parameter:
\begin{equation}
\label{eqn:Qab}
Q_{\alpha\beta}(\mathbf{r},\tau)=\left\{\begin{array}{ccc}\left(\Ket{\alpha}\Bra{\beta}\right)_{\mathbf{r},\tau}-\delta_{\alpha\beta}\frac{\hat{\mathbbold{1}}}{N}&,&\text{A sublattice}\\\left(\Ket{\beta}\Bra{\alpha}\right)_{\mathbf{r},\tau}-\delta_{\alpha\beta}\frac{\hat{\mathbbold{1}}}{N}&,&\text{B sublattice}\end{array}\right.,
\end{equation}
where $\alpha$ and $\beta$ vary over the $N$ colors.  We can then define the zero-frequency magnetic susceptibility as:
\begin{equation}
\label{eqn:chiN}
\chiN\equiv\frac{1}{(N_s\beta)^2}\sum_{\mathbf{r},\mathbf{r}'}\int_0^\beta d\tau\int_0^\beta d\tau'\Braket{\text{T}_\tau Q_{\alpha\beta}(\mathbf{r},\tau)Q_{\beta\alpha}(\mathbf{r}',\tau')}.
\end{equation}
Note that typical definitions of the susceptibility throughout the literature may vary by factors of $N_s\beta$; we choose to divide out this extensiveness in our definition.  Therefore, near a critical point located at $g_c$, the theory of critical phenomena in finite size systems predicts that the susceptibility will fit a form given by
\begin{equation}
\label{eqn:chiNscaled}
\chiN=L^{2-D-\etaN}\mathbb{Y}_{\text{N}}\left[\frac{g-g_c}{g_c}L^{1/\nu}\right],
\end{equation}
where $\mathbb{Y}_{\text{N}}$ is analytic in its argument, $D=2+1$, and $\nu$, the correlation length exponent, and $\etaN$, the anomalous dimension of the N\'{e}el order parameter, are universal critical exponents.  Note that the subscript ``$\text{N}$" on $\etaN$ stands for ``N\'{e}el" and has nothing to do with the $N$ in SU($N$).  Of course, for finite sized systems, there may be sub-leading corrections to this form.  Also, $g$ is the continuously variable coupling, which, in the main text, is either $g=J_2/J_1$
  (honeycomb) and $g=J_2/J^x_1$ or $g=J_1^x/Q^{x,x}$
  (rectangular), but in all cases is distinct and unrelated to the $g$ of Eq.~(\ref{eq:cpN}).

As was done for our model [Eq.~(1) in the main text] on the square lattice ($q=4$), we can ask in what phase we find the ground state for the $J_1$-only model ($J_2=Q^{x,x}=0$) at various integer values of $N$.  On the square lattice, $N=2,3,4$ were found to have N\'{e}el ordered ground states while systems with $N\geq5$ were found to have VBS ordered ground states.  We can check for the presence of magnetic order easily enough and therefore establish that on the honeycomb lattice, we have the N\'{e}el phase again for $N=2,3,4$ and the VBS phase for $N\geq5$ (see Fig.~\ref{fig:chiNJ1only}).  On the rectangular lattice, the situation is more complicated and depends on the anisotropy between $J_{1}^x$ and $J_{1}^y$.  See Sec.~\ref{sec:rectpd} for a detailed discussion of this situation.

The underlying field theory of deconfined quantum criticality is the so-called \cp{N-1} field theory, which has been studied analytically in the limit of large $N$.  The result for $\etaN$, obtained from a $1/N$ expansion of the N\'{e}el order parameter expressed in terms of the \cp{N-1} fields~\cite{kaul2008:u1}, to the highest order currently known is
\begin{equation}
\label{eqn:etaN}
\etaN=1-\frac{32}{\pi^2N}+\ldots
\end{equation}
to which we compare our results in the main text.

\subsection{Binder Ratio}\label{sec:R2}

Starting with the definition of the SU($N$) order parameter in
Eq.~(\ref{eqn:Qab}),  we can define a generalization of the popular
Binder ratio that has been used to identify the location of critical
points (for an introduction, see Ref.~\cite{sandvik2010:vietri}). The main idea is to construct a ratio of two quantities that
have the same scaling dimension, so that the ratio is volume
independent at the critical point. Following Binder's original
suggestion, we construct the ratio, $R_2$, of the average of the fourth power
of the order parameter to the square of the average of the square of
the order parameter. It is natural to contract the indices to maintain
SU($N$) invariance,
\begin{widetext}
\begin{equation}
\label{eqn:R2}
R_2=\frac{\displaystyle\left(\prod_{\mu=1}^4\int_0^\beta d\tau_\mu\sum_{\mathbf{r}_\mu}\right)\Braket{\text{T}_\tau Q_{\alpha\beta}(\mathbf{r}_1,\tau_1)Q_{\beta\alpha}(\mathbf{r}_2,\tau_2)Q_{\gamma\delta}(\mathbf{r}_3,\tau_3)Q_{\delta\gamma}(\mathbf{r}_4,\tau_4)}}{\displaystyle\left[\left(\prod_{\mu=1}^2\int_0^\beta d\tau_\mu\sum_{\mathbf{r}_\mu}\right)\Braket{\text{T}_\tau Q_{\alpha\beta}(\mathbf{r}_1,\tau_1)Q_{\beta\alpha}(\mathbf{r}_2,\tau_2)}\right]^2}.
\end{equation}
\end{widetext}
It is possible to show that this quantity reduces to the familiar
Binder ratio when $N=2$. The virtue of a quantity such as this is
that, as a function of the coupling, $g$, the curves formed by data
sets corresponding to different system sizes should cross at the same
value of $g$, namely $g_c$, without the knowledge of an unknown
parameter ($\etaN$).  Invoking standard finite size scaling arguments,
we expect the Binder ratio to have the following scaling form,
\begin{equation}
\label{eqn:R2scaled}
R_2=\mathbb{Y}_{R_2}\left[\frac{g-g_c}{g_c}L^{1/\nu}\right],
\end{equation}
where $\mathbb{Y}_{R_2}$ is analytic.  As with the susceptibility, there are sub-leading corrections to this scaling form for finite sized systems.

\begin{figure}[h]
\centerline{\includegraphics[width=\columnwidth]{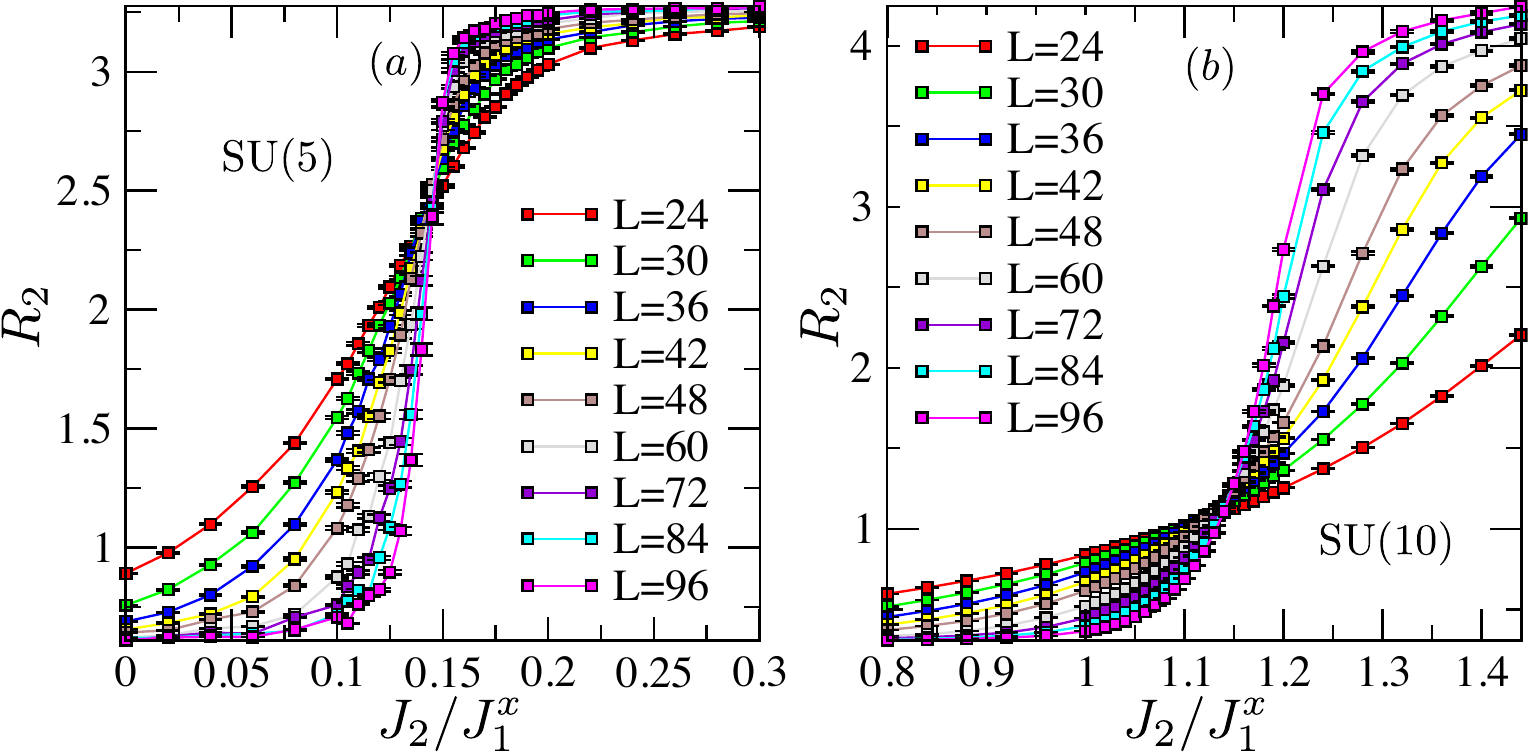}}
\centerline{\includegraphics[width=\columnwidth]{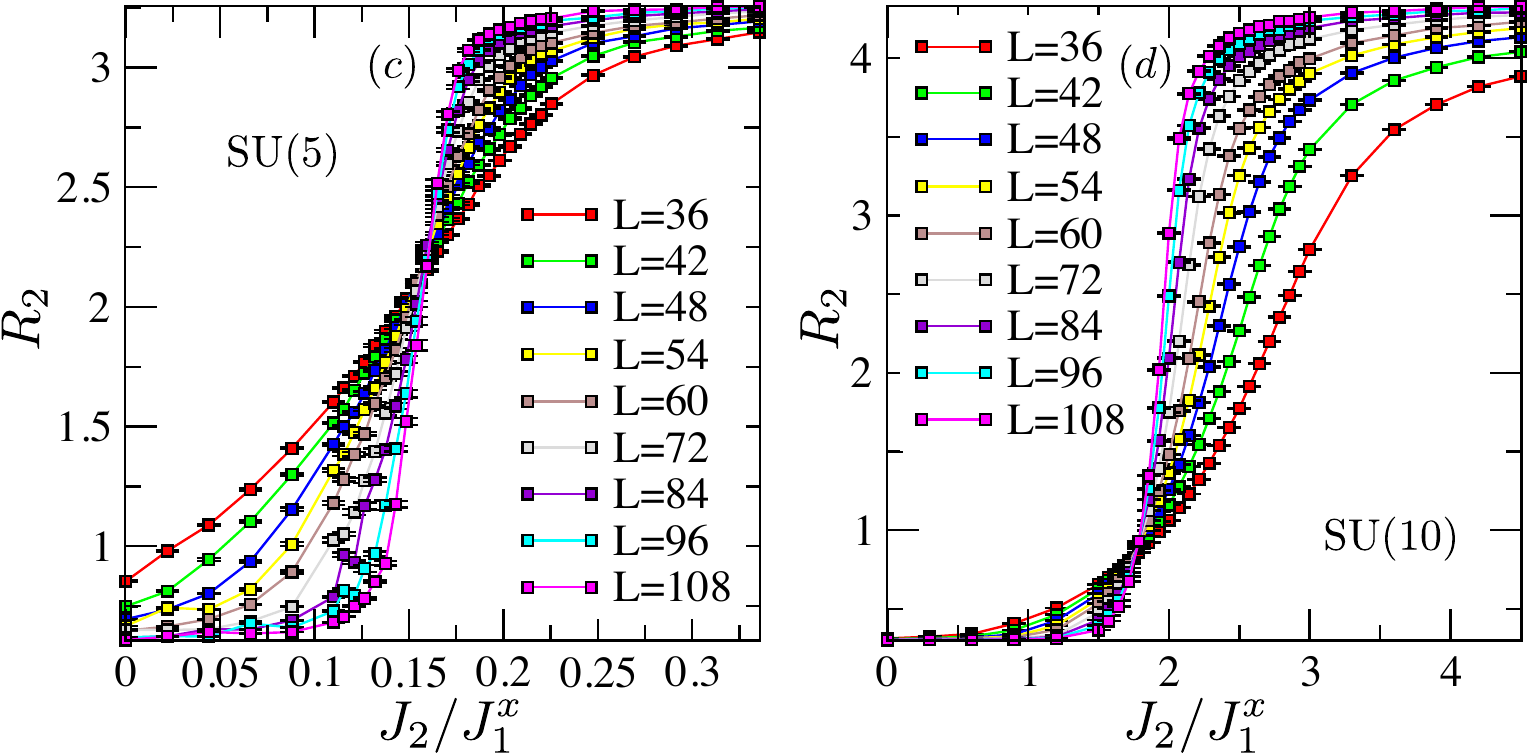}}
\caption{(color online).  Examples of the generalized Binder ratio as
  defined in Eq.~(\ref{eqn:R2}) as a function of $g=J_2/J_1$
  (honeycomb) and $g=J_2/J^x_1$
  (rectangular). (a) SU(5) on the honeycomb lattice; (b) SU(10) on the honeycomb lattice; (c) SU(5) on the rectangular lattice; (d) SU(10) on the rectangular lattice.  The honeycomb lattices used have $2L^2$ sites while the rectangular lattices used have $4L^2/3$ sites.}
\label{fig:R2su5su10}
\end{figure}

In the main text, we show our Binder ratio data for SU(7) on both the honeycomb and rectangular lattices.  Fig.~\ref{fig:R2su5su10} shows the Binder ratio data for SU(5) and SU(10).  Note that this Binder ratio is not normalized in any way, but it is nonetheless clear that the quantity asymptotically approaches two distinct finite values deep within each phase.

\subsection{Spin Stiffness}\label{sec:rho}
\begin{figure}[h]
\centerline{\includegraphics[width=\columnwidth]{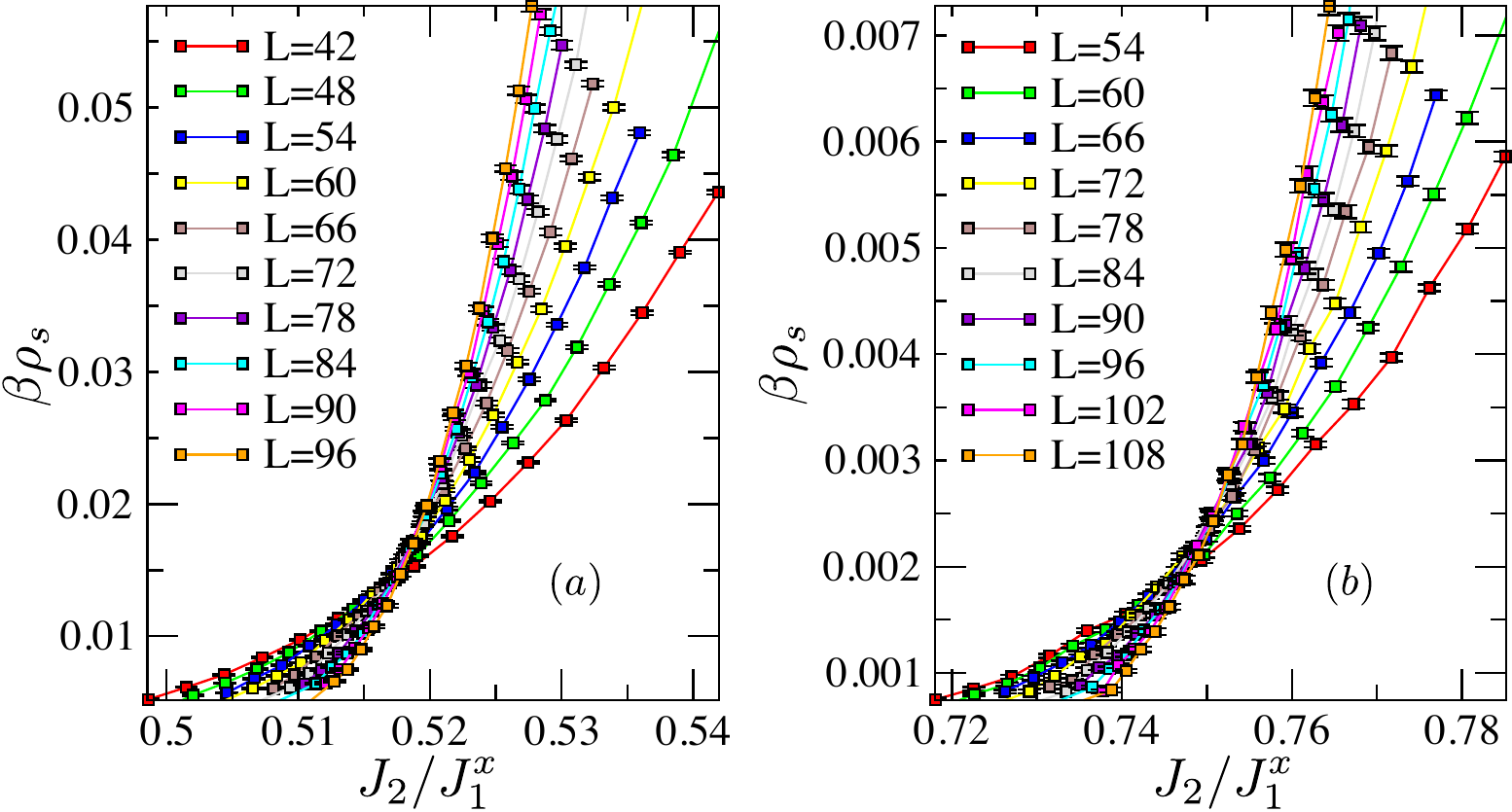}}
\caption{(color online).  Examples of the spin stiffness scaled by $\beta$ ($\beta\rho_s=\Braket{W^2}$) as a function of $g=J_2/J_1$
  (honeycomb) and $g=J_2/J^x_1$
  (rectangular) for SU(7). (a) Honeycomb lattice; (b) rectangular lattice.}
\label{fig:rhos}
\end{figure}

A defining feature of the N\'{e}el phase is a finite spin stiffness $\rho_s$.  In our QMC simulations with global loops updates, we can measure the stiffness very simply by computing the fluctuations of the spatial winding number $W$ of world lines: $\beta\rho_s=\Braket{W^2}$~\cite{sandvik2010:vietri}.  At a point where magnetic fluctuations become critical, the quantity $\beta\rho_s$ becomes $L$-independent; that is to say that it has a scaling form similar to Eq.~(\ref{eqn:R2scaled}) in the vicinity of the critical point, albeit with a different function, $\mathbb{Y}_{\rho}$.  Fig.~\ref{fig:rhos} shows our stiffness data for SU(7) on both the honeycomb and rectangular lattices.

\section{VBS Susceptibility}\label{sec:vbssus}
To determine the presence of the VBS phase, we measure a static ($\omega=0$) VBS susceptibility, $\chiV$.  First we define the bond operator on a pair of nearest neighbor sites as follows:
\begin{equation}
\label{eqn:Bmu}
B^\mu(\mathbf{r},\tau)=\frac{1}{N}\mathcal{P}(\mathbf{r},\tau;\mathbf{r}+\hat{\mu},\tau),
\end{equation}
where $\mathcal{P}$ is the same as that defined in Eq.~(1) in the main text with spacetime locations of the two points given by the arguments.  The superscript $\mu$ denotes the bond type.  On the square or rectangular lattices, this index would run over $\mu=x,y$.  On the honeycomb lattice, there are three distinct bond types with orientations rotated 120$^\circ$ from one another.  We can then study the correlations of these bond operators at different points in space and take the static component:
\begin{widetext}
\begin{equation}
\label{eqn:Cmunu}
C^{\mu\nu}(\mathbf{r}-\mathbf{r}')\equiv\frac{1}{\beta^2}\int_0^\beta d\tau\int_0^\beta d\tau'\Braket{\text{T}_\tau B^\mu(\mathbf{r},\tau)B^\nu(\mathbf{r}',\tau')}-\Braket{B^\mu}\Braket{B^\nu}.
\end{equation}
\end{widetext}
A particular VBS pattern corresponds to a wavevector $\mathbf{Q}$ and correlated bond types $\bar{\mu}$ and $\bar{\nu}$.  For example, on the rectangular lattice where the $J_1$ (Heisenberg) coupling is stronger along the $x$-axis, we expect correlations between $x$-type bonds (and so $\bar{\mu},\bar{\nu}=x$) with wavevector $\mathbf{Q}=(\pi,0)$ (this is a columnar pattern).  By taking the Fourier component of $C^{\bar{\mu}\bar{\nu}}$ at this wavevector, we can check for a signal in this VBS pattern.  This is how we define our VBS susceptibility:
\begin{equation}
\label{eqn:chiV}
\chiV\equiv\frac{1}{N_s}\sum_{\mathbf{r}}C^{\bar{\mu}\bar{\nu}}(\mathbf{r})e^{i\mathbf{Q}\cdot\mathbf{r}}.
\end{equation}

In the vicinity of a critical point, we expect the susceptibility data for different finite size systems to scale as
\begin{equation}
\label{eqn:chiVscaled}
\chiV=L^{2-D-\etaV}\mathbb{Y}_{\text{V}}\left[\frac{g-g_c}{g_c}L^{1/\nu}\right],
\end{equation}
where $D=2+1$, $g_c$ and $\nu$ are expected to have the same values as in the N\'{e}el case, $\mathbb{Y}_{\text{V}}$ is analytic, and $\etaV$, the anomalous dimension of the VBS order parameter, is a new, universal critical exponent.

Analytic work to estimate the value of $\etaV$ was performed~\cite{murthy1990:mono,metlitski2008:mono} by exploiting a nontrivial relation, predating the DQC theory, between monopoles in the field theory and the VBS order on the lattice~\cite{read1989:vbs}. To the highest currently known order
\begin{equation}
\label{eqn:etaV}
\etaV=2\delta_1N-1+\ldots,
\end{equation}
where $\delta_1\approx0.1246$.

\section{Additional Numerical Results}
\begin{figure}[h]
\centerline{\includegraphics[width=\columnwidth]{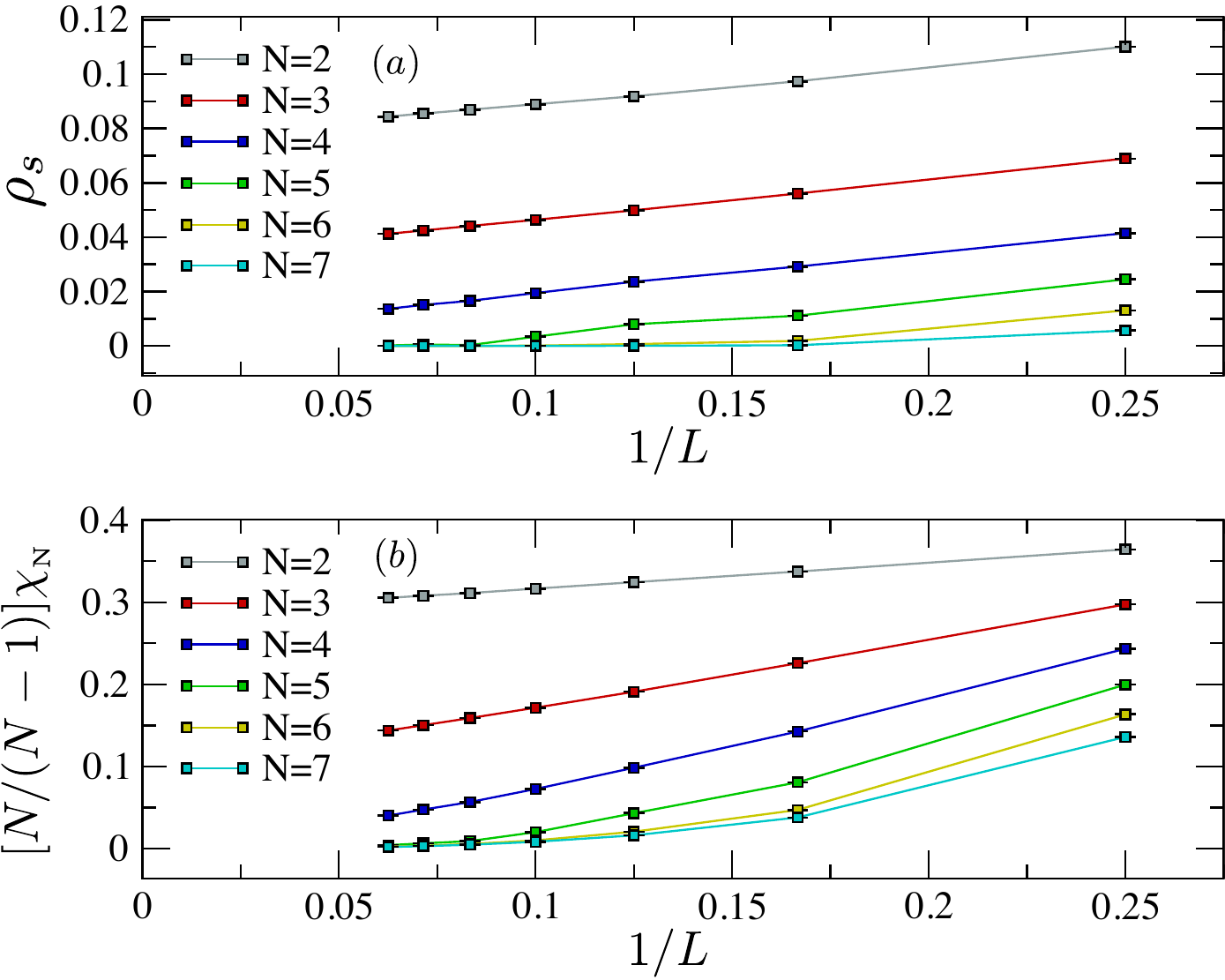}}
\caption{(color online).  (a) Spin stiffness (see Sec.~\ref{sec:rho} for the definition) along one of the three fundamental directions on the honeycomb lattice as a function of $1/L$ for various values of $N$.  (b) (Normalized) Magnetic susceptibility as a function of $1/L$ for various values of $N$. These data demonstrate that a phase transition from N\'{e}el to VBS takes place between $N=4$ and $N=5$.}
\label{fig:chiNJ1only}
\end{figure}

Here we present some additional details concerning our investigation of the phase transitions in the model given by Eq.~(1) in the main text.  First, we consider the presence or absence of magnetic order for the honeycomb lattice version of the model with $J_2=Q^{x,x}=0$ (i.e., the $J_1$-only model) for various values of $N$.  We observe that magnetic order disappears as $N$ is increased from 4 to 5 (Fig.~\ref{fig:chiNJ1only}).

\begin{figure}[h]
\centerline{\includegraphics[width=\columnwidth]{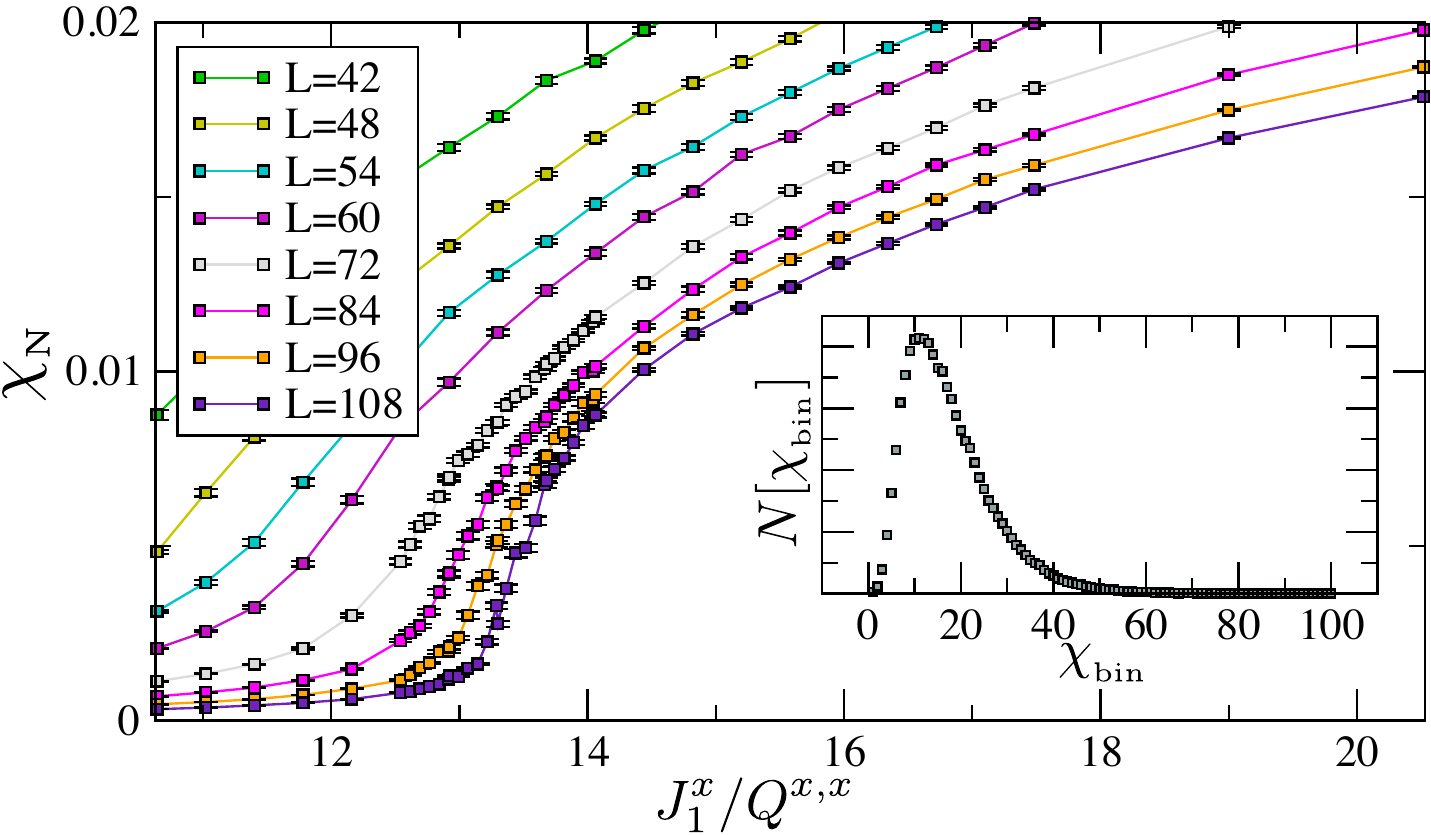}}
\caption{(color online).  Magnetic susceptibility for SU(4) on the rectangular lattice.  Unlike in the SU(3) case, there is no observable jump in the data.  The inset shows a single-peaked histogram of data taken from a point in the middle of the transition ($J_1^x/Q^{x,x}=11.02$) for $L=48$ thus providing further evidence for the nature of the transition.  For this inset, susceptibility data was averaged for 50 measurement sweeps at a time using a total of $8\times10^6$ sweeps.  The averaged values were then placed into 100 equally sized bins spread out over the entire range of observed values.  The ``bin number," $\chi_{_\text{bin}}$, is on the $x$-axis while the number of elements in that bin is shown on the $y$-axis.  The values of $L$ correspond to a rectangular lattice with $4L^2/3$ sites.}
\label{fig:2ndordersu4}
\end{figure}

Next, we consider the phase transitions on the rectangular lattice.  In the main text, we show evidence that the transition for $N=3$ is first-order.  Here, in Fig.~\ref{fig:2ndordersu4} we show that the transition has become continuous for $N=4$.

\begin{figure}[h]
\centerline{\includegraphics[width=\columnwidth]{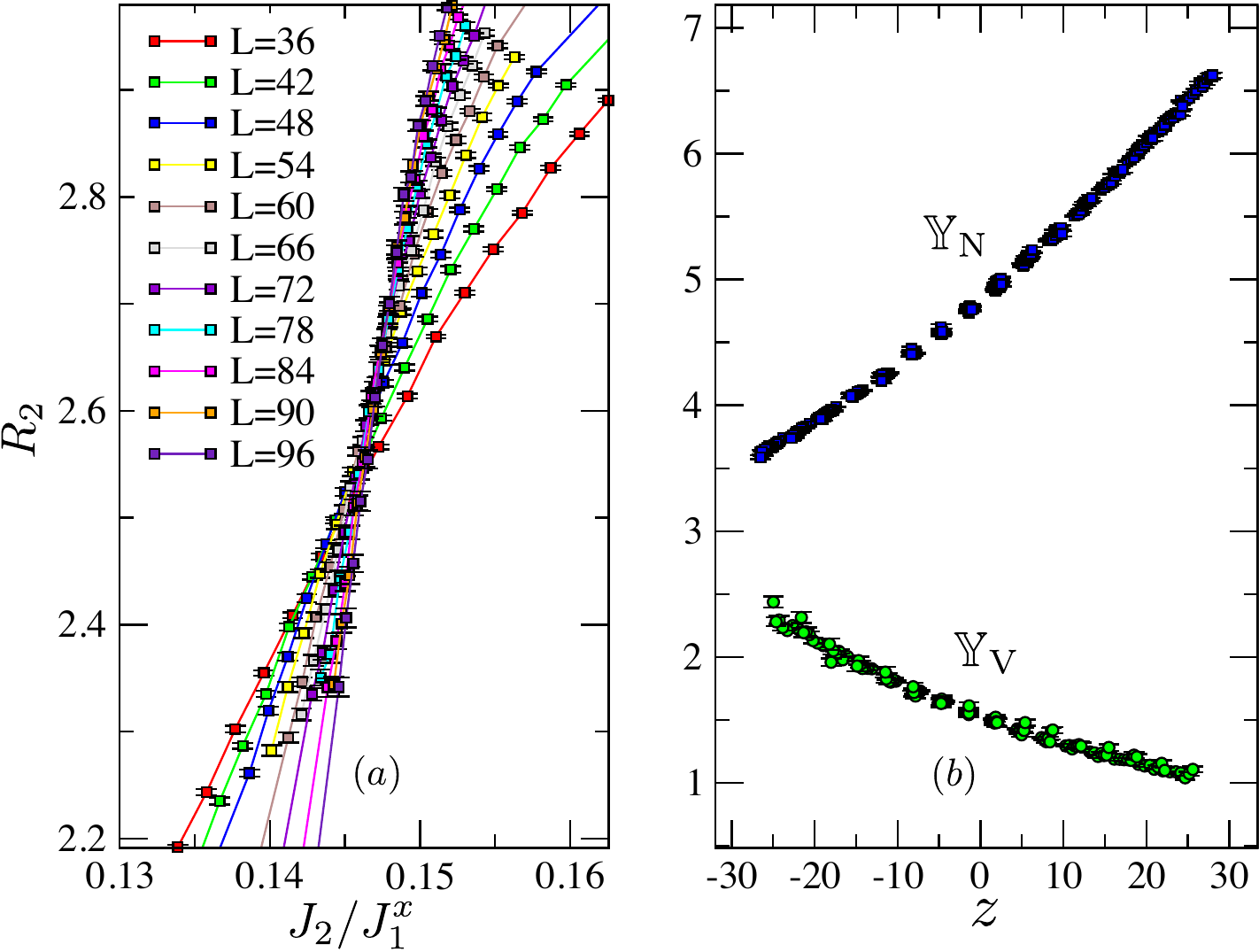}}
\caption{(color online).  This data is for the honeycomb lattice, SU(5). (a) This panel shows the Binder ratio data. (b) Both the magnetic (blue squares) and VBS (green circles) susceptibility data.  The data has been collapsed such that $\mathbb{Y}_{\text{N}}(z)=L^{1+\etaN}\chiN(z)+(a+bz)L^{-\omega}$ and $\mathbb{Y}_{\text{V}}(z)=L^{1+\etaV}\chiV(z)$ with $\etaN=0.646$, $a=23.0$, $b=-0.125$, $\omega=1.0$, and $\etaV=1.06$.  Also, $g_c=0.1481$ and $\nu=0.65$ for the purpose of converting $g=J_2/J_1$
  (honeycomb) and $g=J_2/J^x_1$
  (rectangular) to $z$.  For the magnetic susceptibility, the following system sizes were used in the collapse: $L=36,42,48,54,60,66,72,78,84,90,96$.  For the VBS susceptibility, the following system sizes were used in the collapse: $L=18,24,30,36,42,48,54$.  There are $2L^2$ lattice sites.}
\label{fig:hcSU5_r2_sus}
\end{figure}

\begin{figure}[h]
\centerline{\includegraphics[width=\columnwidth]{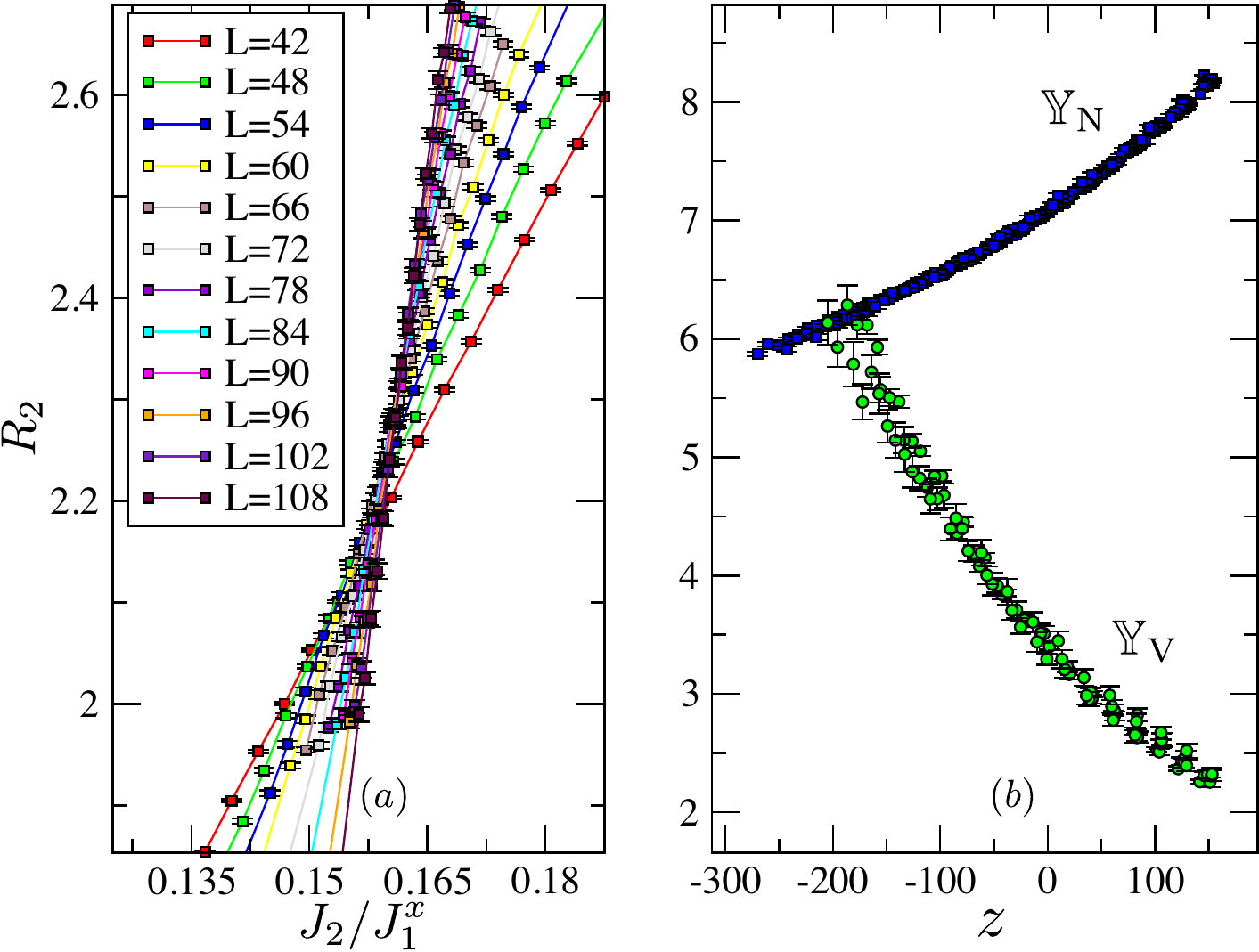}}
\caption{(color online).  This data is for the rectangular lattice, SU(5). (a) This panel shows the Binder ratio data. (b) Both the magnetic (blue squares) and VBS (green circles) susceptibility data.  The data has been collapsed such that $\mathbb{Y}_{\text{N}}(z)=L^{1+\etaN}\chiN(z)+(a+bz)L^{-\omega}$ and $\mathbb{Y}_{\text{V}}(z)=L^{1+\etaV}\chiV(z)$ with $\etaN=0.599$, $a=10.8$, $b=-0.028$, $\omega=0.5$, and $\etaV=0.679$.  Also, $g_c=0.1639$ and $\nu=0.54$ for the purpose of converting $g=J_2/J_1$
  (honeycomb) and $g=J_2/J^x_1$
  (rectangular) to $z$.  For the magnetic susceptibility, the following system sizes were used in the collapse: $L=42,48,54,60,66,72,78,84,90,96,102,108$.  For the VBS susceptibility, the following system sizes were used in the collapse: $L=36,42,48,54,60,66$. There are $4L^2/3$ lattice sites.}
\label{fig:anrecSU5_r2_sus}
\end{figure}

\begin{figure}[h]
\centerline{\includegraphics[width=\columnwidth]{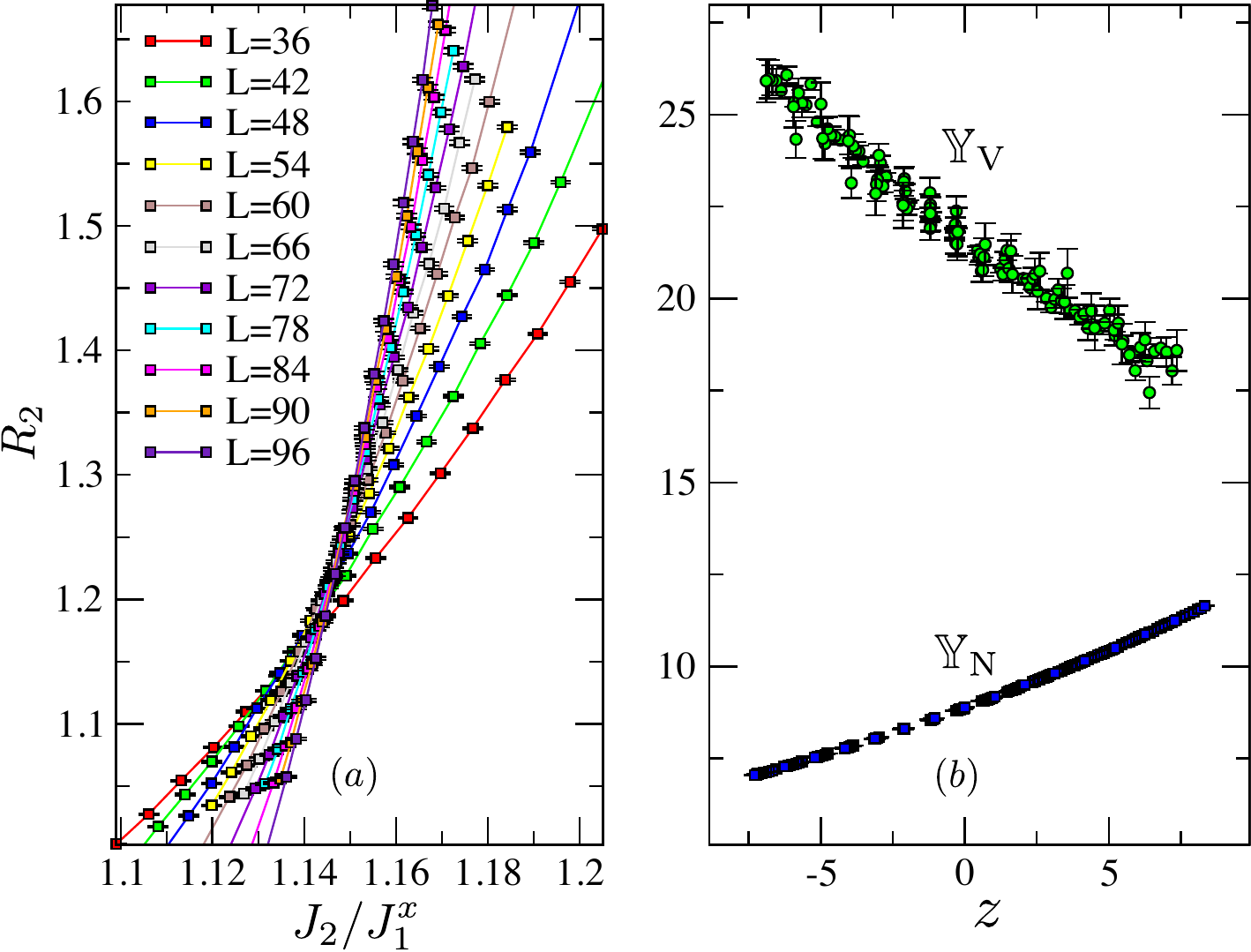}}
\caption{(color online).  This data is for the honeycomb lattice, SU(10). (a) This panel shows the Binder ratio data. (b) Both the magnetic (blue squares) and VBS (green circles) susceptibility data.  The data has been collapsed such that $\mathbb{Y}_{\text{N}}(z)=L^{1+\etaN}\chiN(z)+(a+bz)L^{-\omega}$ and $\mathbb{Y}_{\text{V}}(z)=L^{1+\etaV}\chiV(z)$ with $\etaN=0.76$, $a=46.5$, $b=1.0$, $\omega=1.0$, and $\etaV=1.71$.  Also, $g_c=1.151$ and $\nu=0.72$ for the purpose of converting $g=J_2/J_1$
  (honeycomb) and $g=J_2/J^x_1$
  (rectangular) to $z$.  For the magnetic susceptibility, the following system sizes were used in the collapse: $L=36,42,48,54,60,66,72,78,84,90,96$.  For the VBS susceptibility, the following system sizes were used in the collapse: $L=18,24,30,36,42,48,54$.  There are $2L^2$ lattice sites.}
\label{fig:hcSU10_r2_sus}
\end{figure}

\begin{figure}[h]
\centerline{\includegraphics[width=\columnwidth]{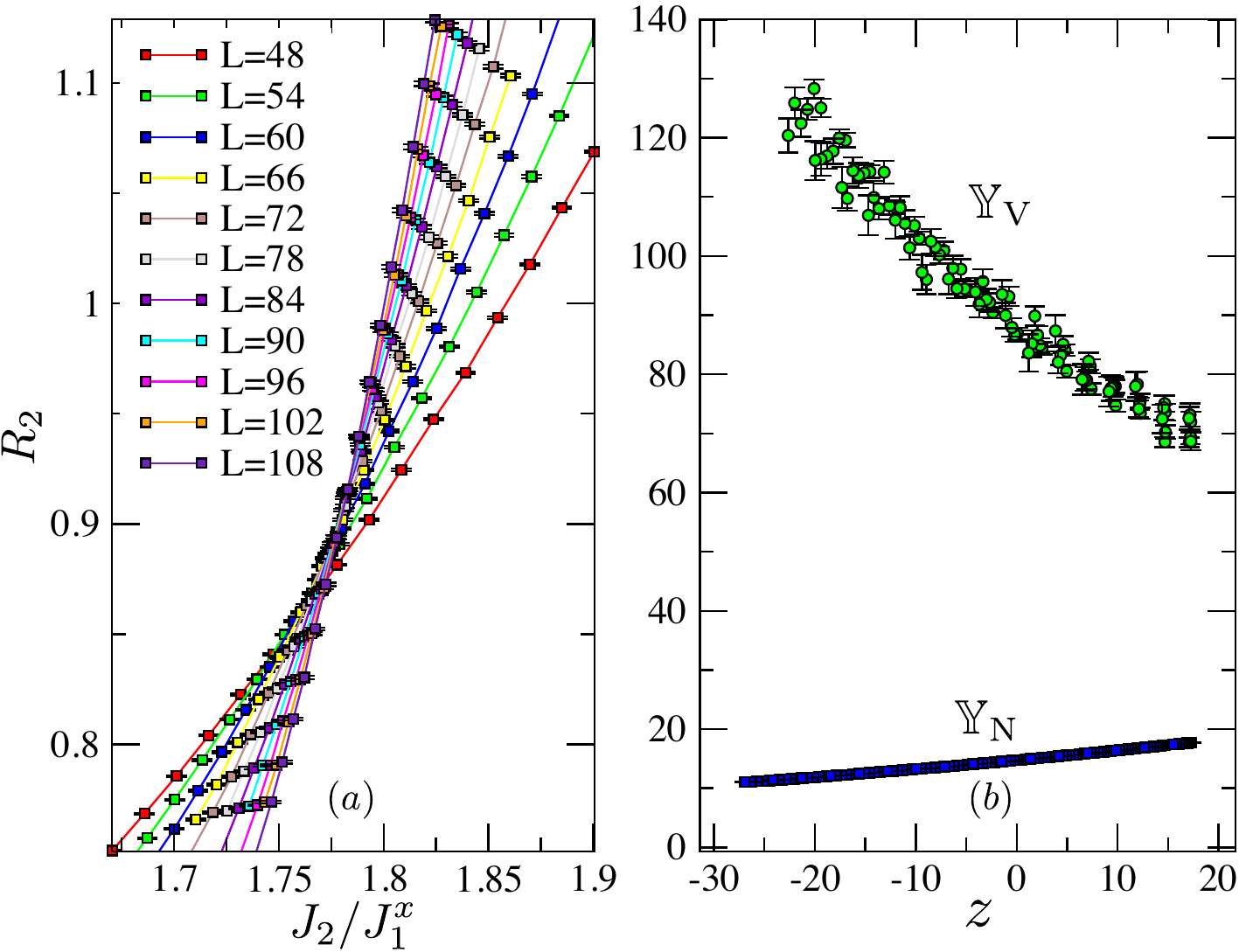}}
\caption{(color online).  This data is for the rectangular lattice, SU(10). (a) This panel shows the Binder ratio data. (b) Both the magnetic (blue squares) and VBS (green circles) susceptibility data.  The data has been collapsed such that $\mathbb{Y}_{\text{N}}(z)=L^{1+\etaN}\chiN(z)+(a+bz)L^{-\omega}$ and $\mathbb{Y}_{\text{V}}(z)=L^{1+\etaV}\chiV(z)$ with $\etaN=0.75$, $a=28.0$, $b=0.2$, $\omega=0.5$, and $\etaV=1.61$.  Also, $g_c=1.796$ and $\nu=0.68$ for the purpose of converting $g=J_2/J_1$
  (honeycomb) and $g=J_2/J^x_1$
  (rectangular) to $z$.  For the magnetic susceptibility, the following system sizes were used in the collapse: $L=42,48,54,60,66,72,78,84,90,96,102,108$.  For the VBS susceptibility, the following system sizes were used in the collapse: $L=36,42,48,54,60,66$. There are $4L^2/3$ lattice sites.}
\label{fig:anrecSU10_r2_sus}
\end{figure}

In the main text, we showed Binder ratio and collapsed susceptibility results for SU(7) on both lattices.  Here, we show the same results for SU(5) in Figs.~\ref{fig:hcSU5_r2_sus}~and~\ref{fig:anrecSU5_r2_sus} and for SU(10) in Figs.~\ref{fig:hcSU10_r2_sus}~and~\ref{fig:anrecSU10_r2_sus}.

\section{Phase Diagram of Rectangular Model}\label{sec:rectpd}
\begin{figure}[h]
\centerline{\includegraphics[width=\columnwidth]{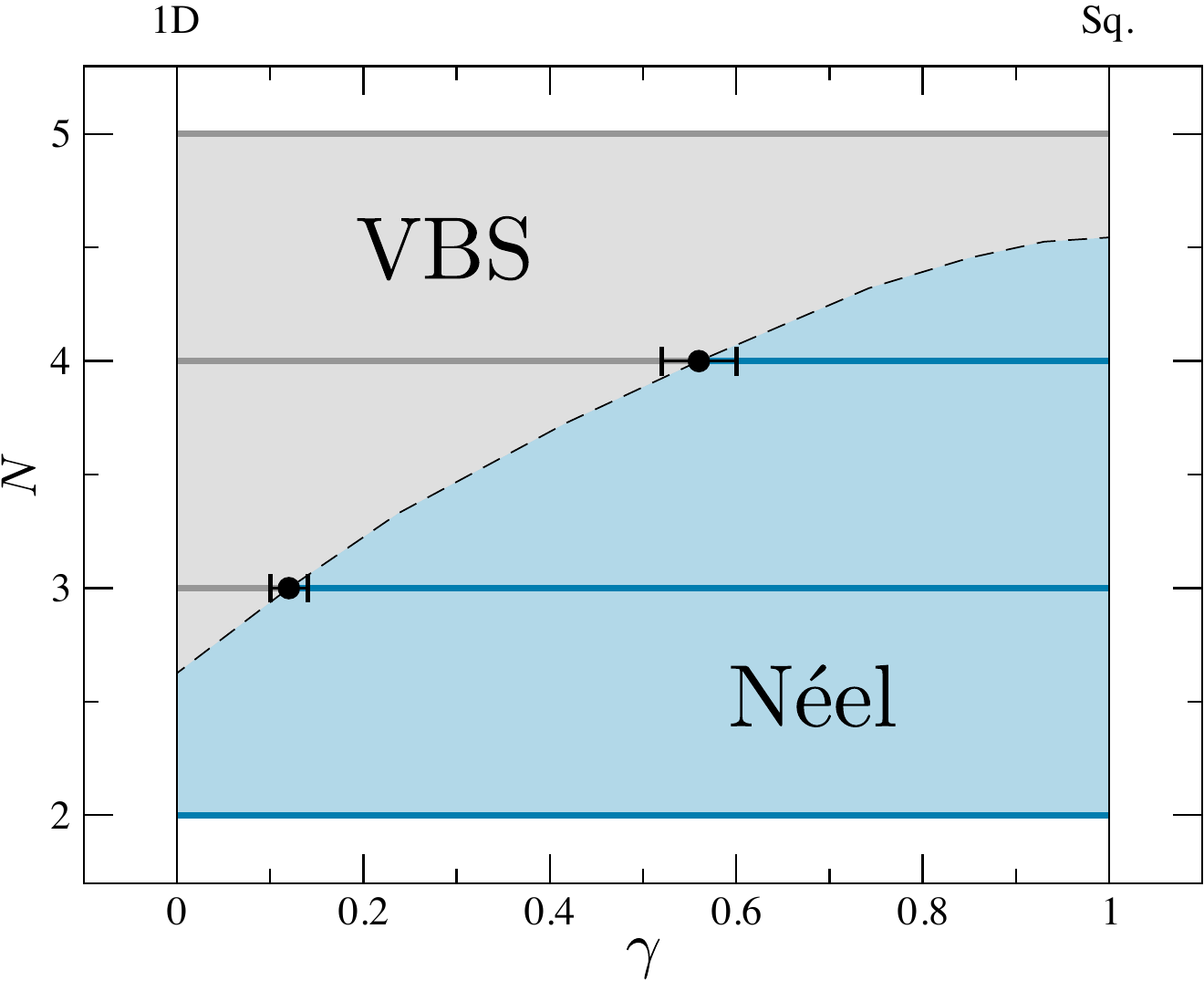}}
\caption{(color online).  The horizontal axis shows the anisotropy in the $J_1$ coupling between the $x$ and $y$ directions on the rectangular lattice such that $J_1^y=\gamma J_1^x$. The vertical axis shows the relevant values of $N$. The black circles indicate the estimated values of $\gamma$ at which the system transitions between N\'{e}el and VBS ordered phases. The dotted line is merely a schematic phase boundary.  Note that SU(2) is always N\'{e}el ordered and SU(5) is always VBS ordered in the $J_1$-only model.}
\label{fig:Ncrit}
\end{figure}

Here we consider the $J_1$-only model on the rectangular lattice, but unlike in the main text where the anisotropy is fixed ($J_1^y=0.8J_1^x$) we instead allow the anisotropy, $\gamma$, to vary as a parameter in the model such that $J_1^y=\gamma J_1^x$.  Studies of this model on a one-dimensional (1D) chain, which corresponds to $\gamma=0$ here, have shown that the SU(2) version is in the so-called Bethe phase (the 1D analog of the N\'{e}el phase) while already for SU(3) the system acquires VBS order.~\cite{affleck1985:lgN,barber1989:d1n3_vbs} Meanwhile, investigation of the square lattice case ($\gamma=1$) has shown that the N\'{e}el-VBS transition occurs somewhere between $N=4$ and $N=5$.~\cite{harada2003:sun,beach2009:sun} Hence, it is reasonable to assume that for $N=3,4$, there exists some finite value of $\gamma<1$ for which the ground state transitions between N\'{e}el and VBS ordered phases.  By considering the Binder ratio (see Sec.~\ref{sec:R2}) for a range of values of $\gamma$ and for a series of system sizes, $N_s=L_xL_y=128,512,2048,8192$, we were able to estimate the location of the transition and visually determine reasonable error bars for our estimate.  There is some subtlety required in choosing appropriate aspect ratios for the geometry of the system, especially when $\gamma$ is small.  Since this aspect ratio varies, we describe the system sizes in terms of number of sites, $N_s$, rather than linear dimensions. The results of our analysis are shown in Fig.~\ref{fig:Ncrit}.

\section{Analysis of Critical Properties}\label{sec:critprop}
The estimation of the location of continuous critical points in the thermodynamic limit as well as the extraction of various critical exponents is a very delicate and challenging endeavor.  While the data for various quantities described in Secs.~\ref{sec:magquant}~and~\ref{sec:vbssus} above for different system sizes should collapse neatly to a single analytic function for each quantity, the reality is that there can be significant, $L$-dependent, sub-leading corrections to scaling and accounting for these (or failing to) can dramatically impact the estimates of various critical quantities.  Indeed, two researchers studying the same data would likely arrive at somewhat different results depending on the method; that is to say, the systematic error of any procedure is assumed to be large.

Throughout the discussion in this section, it should be noted that we typically have very precise data for all of the magnetic quantities of interest.  The error bars (corresponding to stochastic error in the Monte Carlo) are often too small to be visible.  The data for the VBS susceptibility, on the other hand, is quite a bit noisier despite coming from the same number of measurement sweeps.  This is a consequence of the nature of our algorithm, which excels at sampling the magnetic phases efficiently, but slows considerably in the VBS ordered phases.  Nonetheless, our VBS data, especially for the smaller system sizes, is suitably well converged to give meaningful information about the anomalous dimension of the VBS order parameter, $\etaV$.  Ideally, we would obtain more data to increase the precision to the level of the magnetic data; doing so, however, would not add substantially to our main conclusion, namely that the $q=2$ (rectangular), $q=3$ (honeycomb), and $q=4$ (square) versions of our model for $N\geq5$ belong to the same universality class.  The limitations of particular data sets will be addressed specifically below.

\subsection*{Collapse of Data Within Critical Regime}\label{sec:collapse}
\begin{figure}[h]
\centerline{\includegraphics[width=\columnwidth]{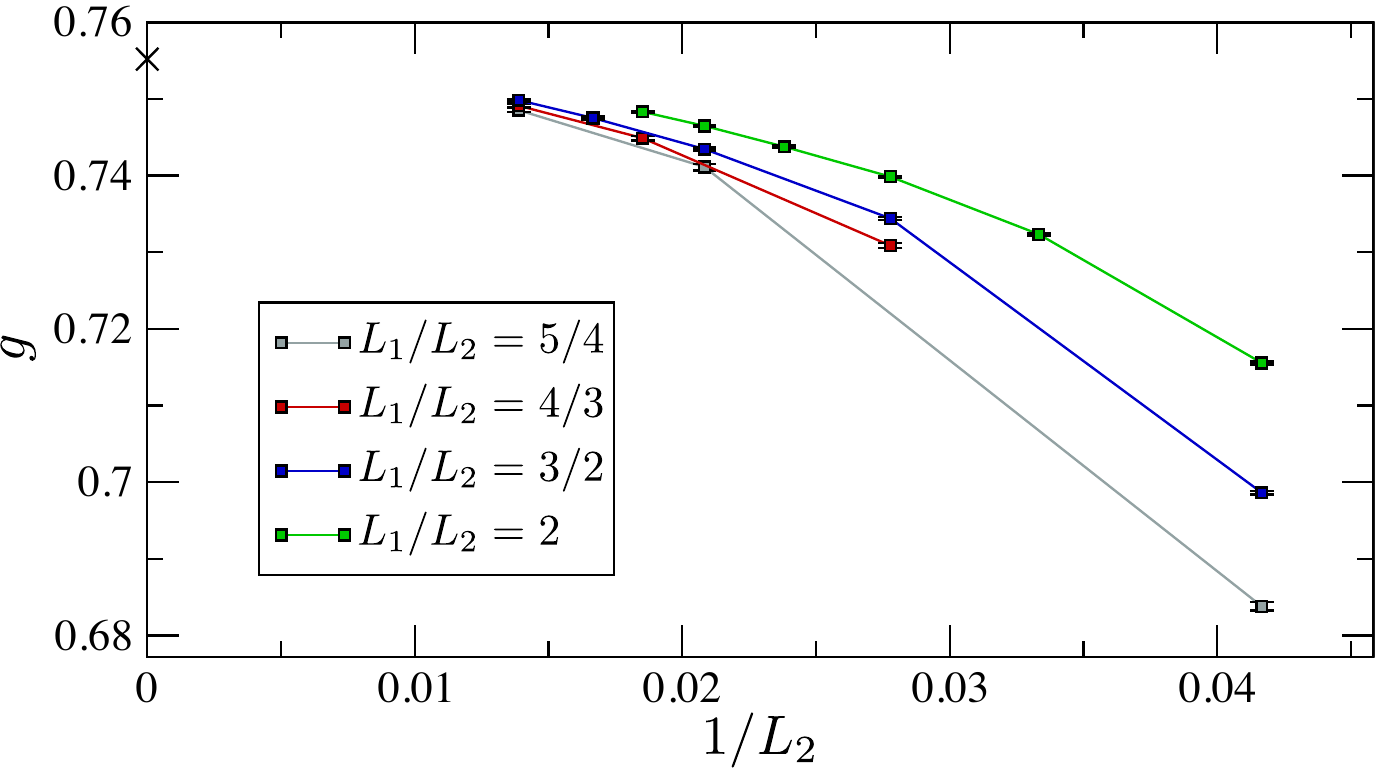}}
\caption{(color online).  Here we estimate the location of the crossing of interpolated curves fitted to the Binder ratio data for SU(7) on the rectangular lattice for pairs of system sizes with various ratios between them.  Each ratio generates a series of crossing locations, $g$, that are then plotted as a function of $1/L$.  By extrapolating the curves to the vertical axis, we can predict a window within which we expect the critical coupling, $g_c$, to live. The point marked with a $\times$ on the vertical axis indicates the value eventually chosen for curve collapse at a later stage in the critical analysis.}
\label{fig:crossingsanrecSU7}
\end{figure}

The locations of the crossings between Binder ratio data curves of different system sizes can be used to estimate a window of values of the coupling $g$ within which we expect the location of the critical point in the thermodynamic limit, $g_c$, to live (see, for example, Fig.~\ref{fig:crossingsanrecSU7}).  This allows us to zoom in on the critical region and collect data near $g_c$ for the purpose of critical collapse (see Sec.~\ref{sec:collapse}).

Once the critical region is identified with sufficient precision, accomplished by iteratively zooming in and analyzing the Binder ratio data, a rough estimate of $g_c$ and $\nu$ can be obtained by attempting to collapse the three magnetic quantities (Binder ratio, magnetic susceptibility, and spin stiffness) to the scaling forms indicated in Sec.~\ref{sec:magquant}.  One of the difficulties in doing such collapses is that ideally one needs data for each system size over a range of values of $z=\left[\left(g-g_c\right)L^{1/\nu}/g_c\right]$ in the critical region.  Choosing values of $g$ that are the same for different lengths will result in the data for larger system sizes spanning a greater space in terms of $z$ than the data for smaller system sizes.  But by using the rough estimates of $g_c$ and $\nu$ from earlier data sets, and specifying a well-defined range of values of $z$, one can generate data sets for different values of $L$ that will be spread out in $z$ (i.e., inverting the definition of $z$ to solve for $g$; e.g., choose 16 equally spaced points on the domain $-10<z<10$ and find the corresponding values of $g$ for each system size given a guess for $g_c$ and $\nu$).  The result is that the values of $g$ will be very different for each system size with smaller system sizes spanning larger regions of $g$ space.  This makes sense in the context of critical phenomena wherein the effective critical region is larger for smaller system sizes and only converges to a point in the thermodynamic limit.  In the main text, the left panels of Figs.~3~and~4 show the uncollapsed Binder ratio data and the varying ranges of $g$ for different system sizes is readily apparent.

\begin{figure}[h]
\centerline{\includegraphics[width=\columnwidth]{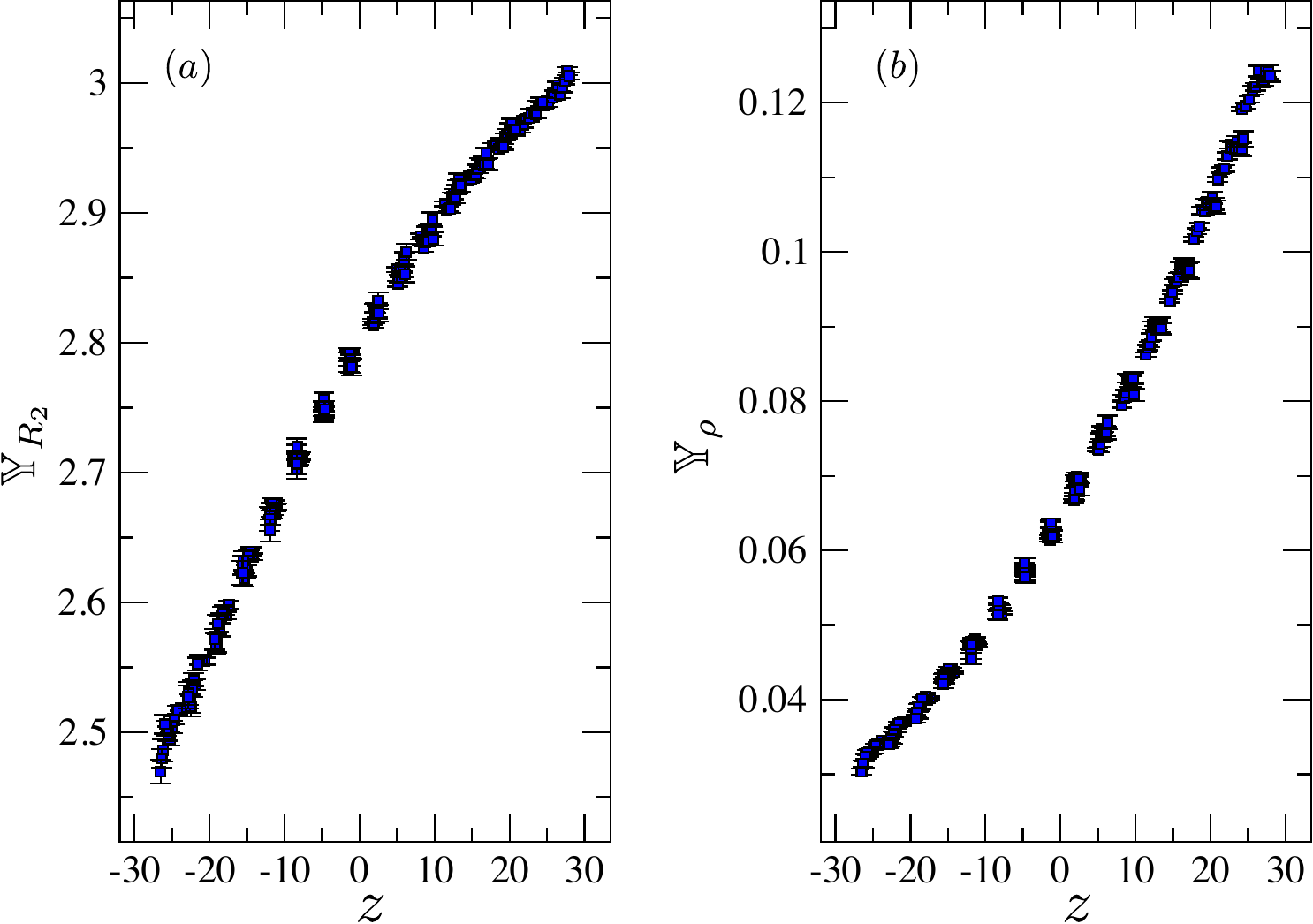}}
\caption{(color online).  Data for honeycomb lattice, SU(5). (a) Binder ratio where $\mathbb{Y}_{R_2}(z)=R_2(z)+(a+bz)L^{-\omega}$ with $a=7.6$, $b=-0.175$, and $\omega=1.0$. (b) The inverse temperature times the spin stiffness where $\mathbb{Y}_{\rho}(z)=\beta\rho_s(z)+(a+bz)L^{-\omega}$ with $a=0.485$, $b=-0.0015$, and $\omega=1.0$.  In both panels, the values $g_c=0.1481$ and $\nu=0.65$ are used to define $z$ and data from the following system sizes are included: $L=36,42,48,54,60,66,72,78,84,90,96$.  The lattices have $2L^2$ sites.}
\label{fig:r2rho_hcSU5}
\end{figure}

\begin{figure}[h]
\centerline{\includegraphics[width=\columnwidth]{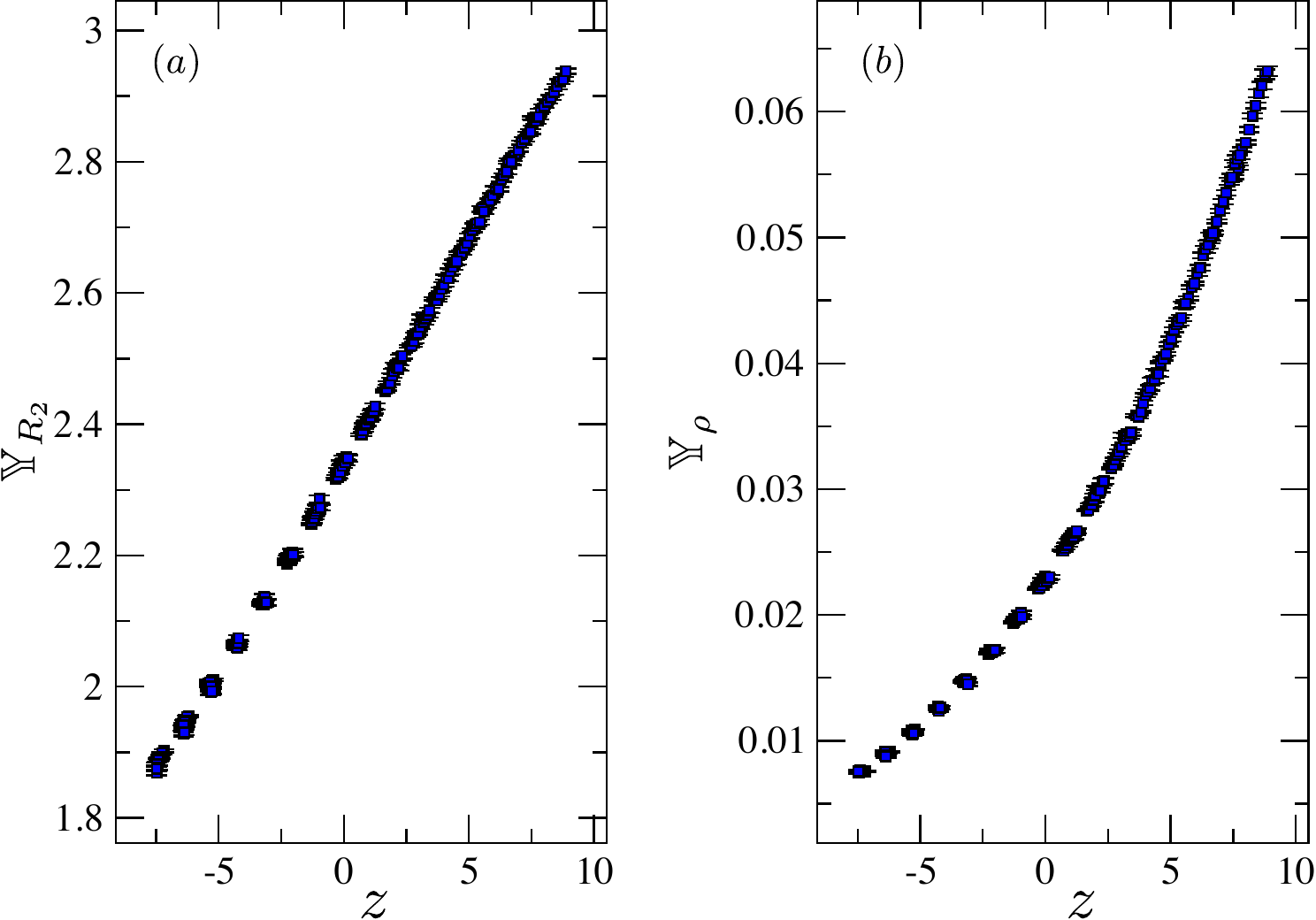}}
\caption{(color online).  Data for honeycomb lattice, SU(7). (a) Binder ratio where $\mathbb{Y}_{R_2}(z)=R_2(z)+(a+bz)L^{-\omega}$ with $a=8.5$, $b=0.0$, and $\omega=1.0$. (b) The inverse temperature times the spin stiffness where $\mathbb{Y}_{\rho}(z)=\beta\rho_s(z)+(a+bz)L^{-\omega}$ with $a=0.295$, $b=0.027$, and $\omega=1.0$.  In both panels, the values $g_c=0.5196$ and $\nu=0.72$ are used to define $z$ and data from the following system sizes are included: $L=36,42,48,54,60,66,72,78,84,90,96$.  The lattices have $2L^2$ sites.}
\label{fig:r2rho_hcSU7}
\end{figure}

\begin{figure}[h]
\centerline{\includegraphics[width=\columnwidth]{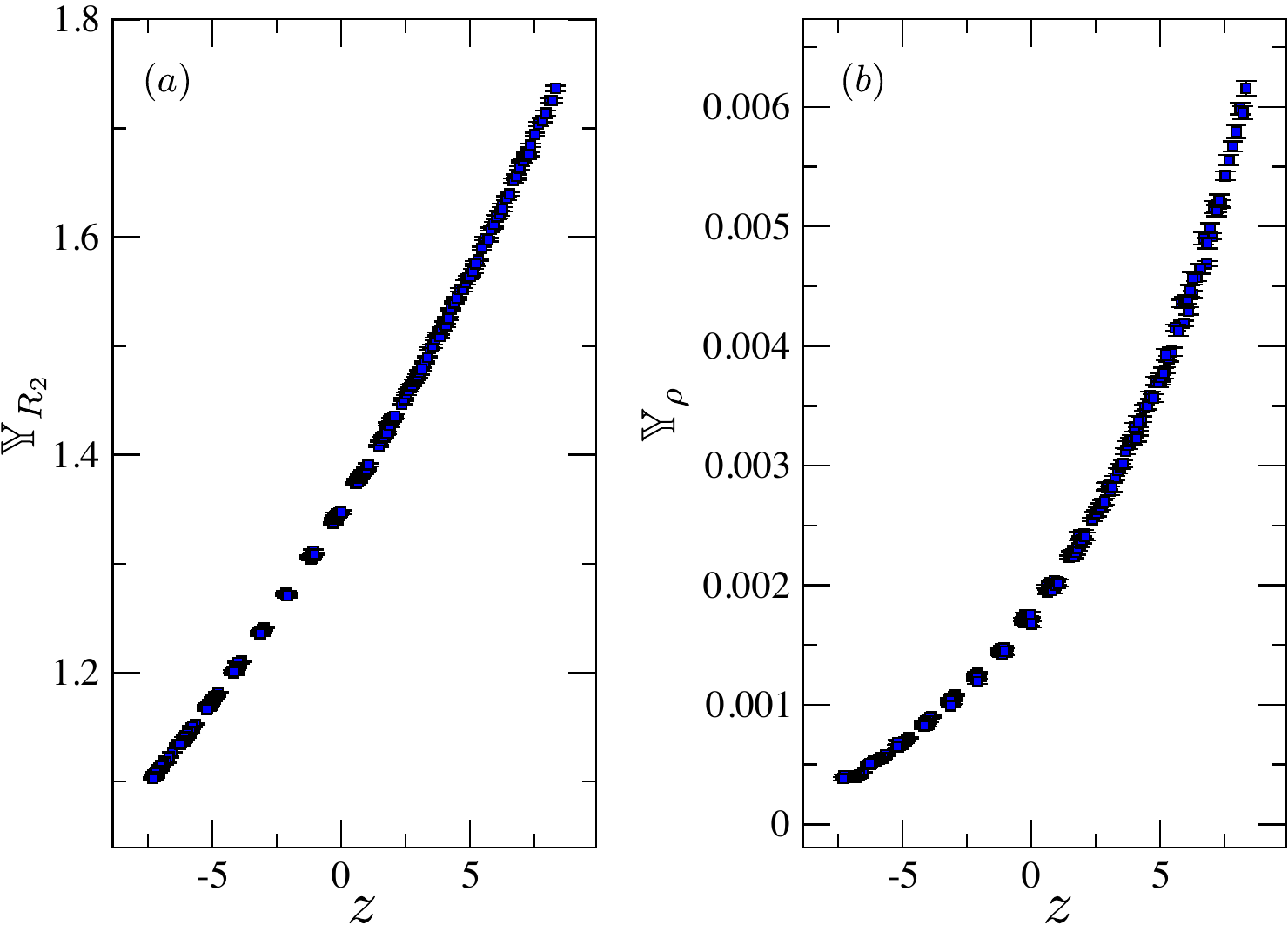}}
\caption{(color online).  Data for honeycomb lattice, SU(10). (a) Binder ratio where $\mathbb{Y}_{R_2}(z)=R_2(z)+(a+bz)L^{-\omega}$ with $a=5.0$, $b=0.09$, and $\omega=1.0$. (b) The inverse temperature times the spin stiffness where $\mathbb{Y}_{\rho}(z)=\beta\rho_s(z)+(a+bz)L^{-\omega}$ with $a=0.0355$, $b=0.00425$, and $\omega=1.0$.  In both panels, the values $g_c=1.151$ and $\nu=0.72$ are used to define $z$ and data from the following system sizes are included: $L=36,42,48,54,60,66,72,78,84,90,96$.  The lattices have $2L^2$ sites.}
\label{fig:r2rho_hcSU10}
\end{figure}

\begin{figure}[h]
\centerline{\includegraphics[width=\columnwidth]{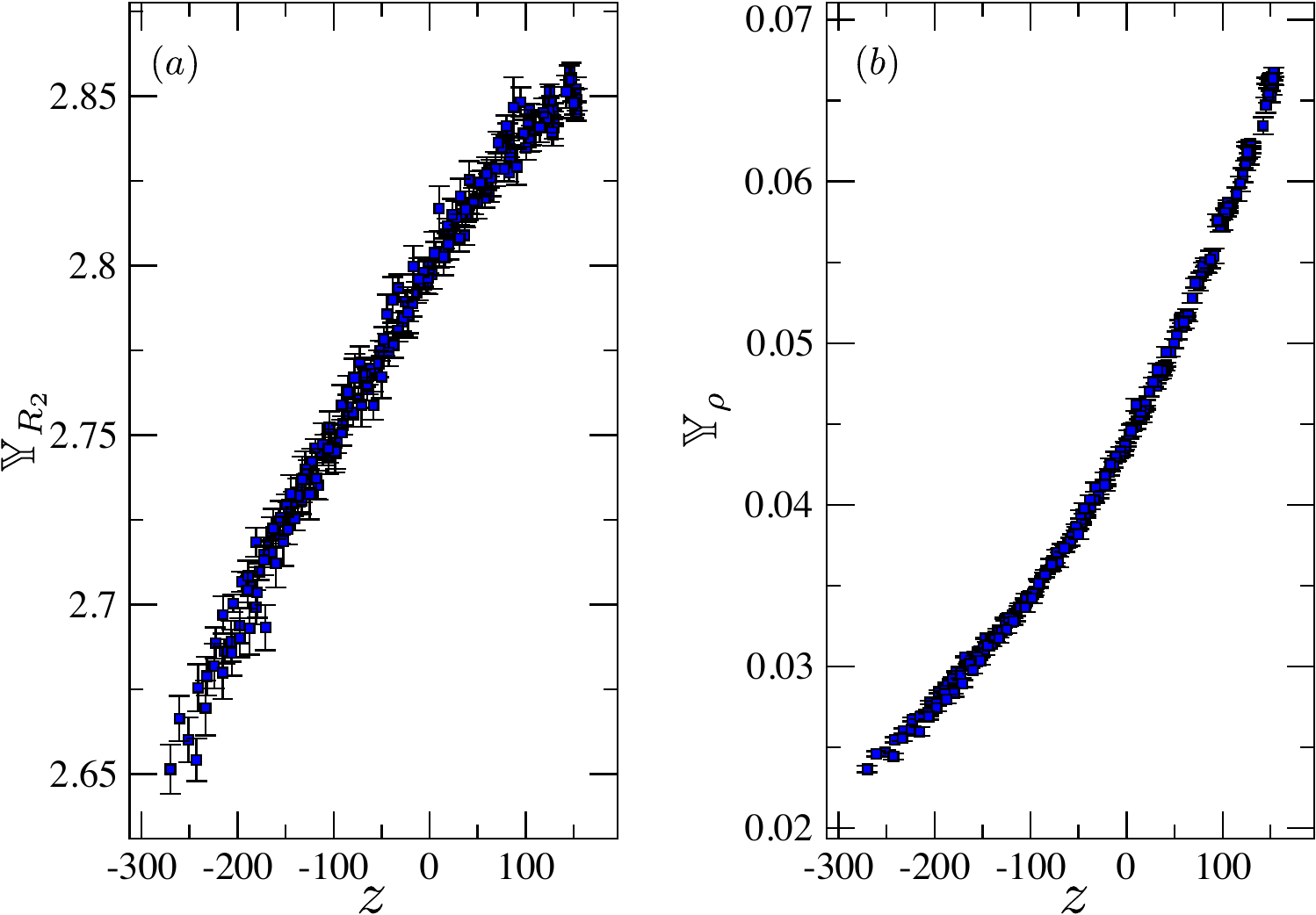}}
\caption{(color online).  Data for rectangular lattice, SU(5). (a) Binder ratio where $\mathbb{Y}_{R_2}(z)=R_2(z)+(a+bz)L^{-\omega}$ with $a=3.5$, $b=-0.0125$, and $\omega=0.5$. (b) The inverse temperature times the spin stiffness where $\mathbb{Y}_{\rho}(z)=\beta\rho_s(z)+(a+bz)L^{-\omega}$ with $a=0.15$, $b=0$, and $\omega=0.5$.  In both panels, the values $g_c=0.1639$ and $\nu=0.54$ are used to define $z$ and data from the following system sizes are included: $L=42,48,54,60,66,72,78,84,90,96,102,108$.  The lattices have $4L^2/3$ sites.}
\label{fig:r2rho_anrecSU5}
\end{figure}

\begin{figure}[h]
\centerline{\includegraphics[width=\columnwidth]{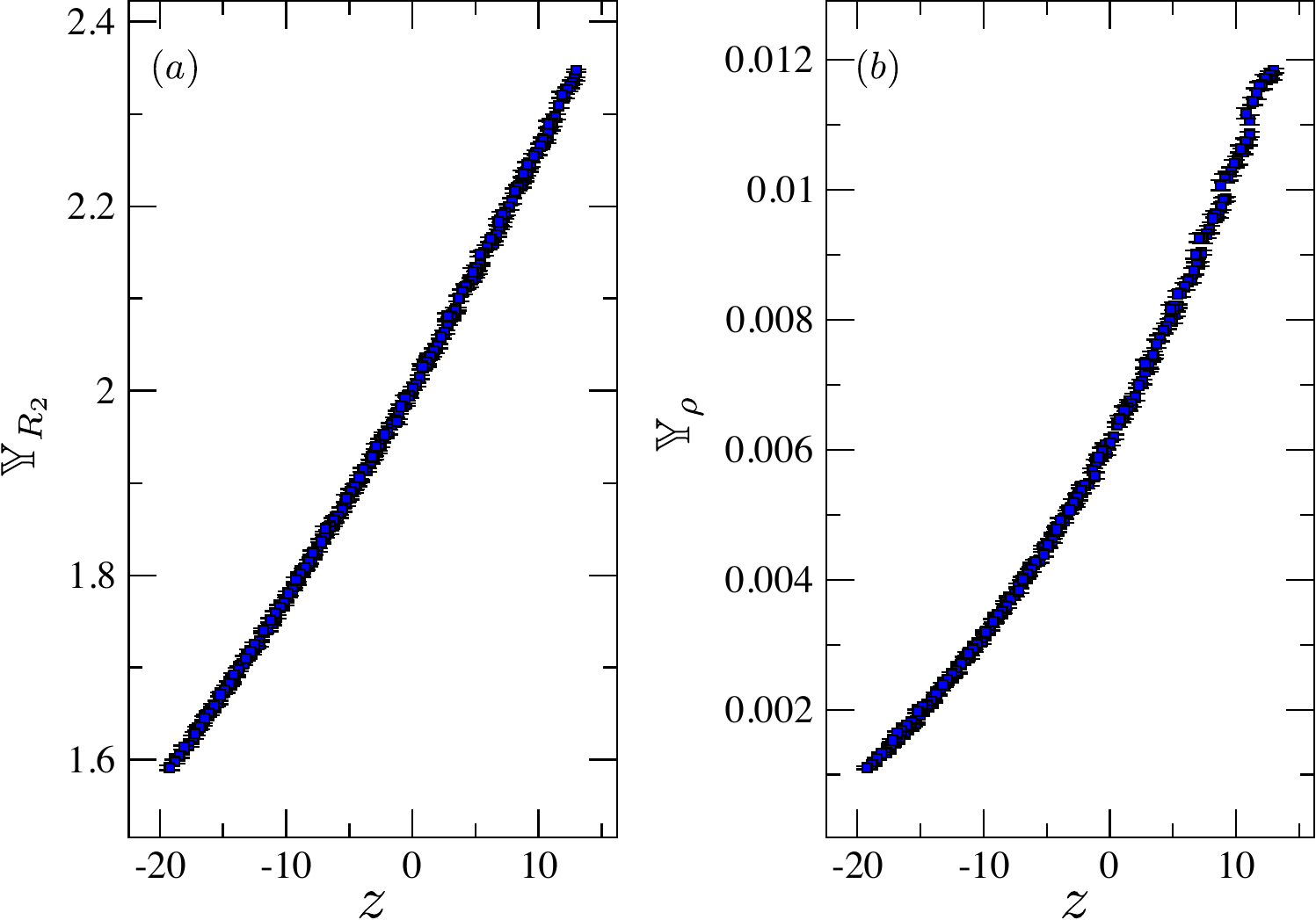}}
\caption{(color online).  Data for rectangular lattice, SU(7). (a) Binder ratio where $\mathbb{Y}_{R_2}(z)=R_2(z)+(a+bz)L^{-\omega}$ with $a=2.64$, $b=0.0224$, and $\omega=0.5$. (b) The inverse temperature times the spin stiffness where $\mathbb{Y}_{\rho}(z)=\beta\rho_s(z)+(a+bz)L^{-\omega}$ with $a=0.027$, $b=0.00126$, and $\omega=0.5$.  In both panels, the values $g_c=0.7552$ and $\nu=0.69$ are used to define $z$ and data from the following system sizes are included: $L=42,48,54,60,66,72,78,84,90,96,102,108$.  The lattices have $4L^2/3$ sites.}
\label{fig:r2rho_anrecSU7}
\end{figure}

\begin{figure}[h]
\centerline{\includegraphics[width=\columnwidth]{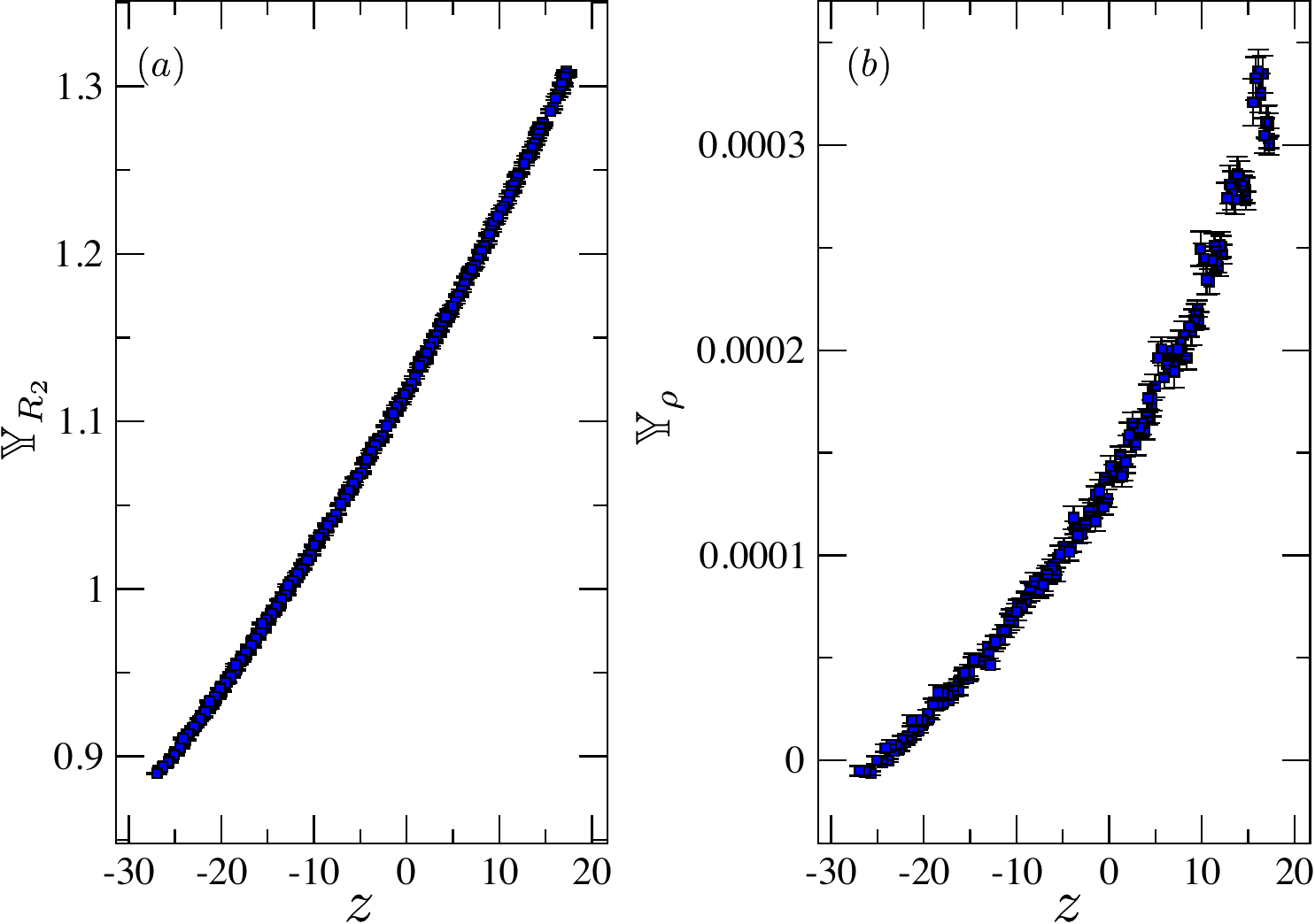}}
\caption{(color online).  Data for rectangular lattice, SU(10). (a) Binder ratio where $\mathbb{Y}_{R_2}(z)=R_2(z)+(a+bz)L^{-\omega}$ with $a=1.474$, $b=0.01$, and $\omega=0.5$. (b) The inverse temperature times the spin stiffness where $\mathbb{Y}_{\rho}(z)=\beta\rho_s(z)+(a+bz)L^{-\omega}$ with $a=7.70\times10^{-4}$, $b=3.36\times10^{-5}$, and $\omega=0.5$.  In both panels, the values $g_c=1.796$ and $\nu=0.68$ are used to define $z$ and data from the following system sizes are included: $L=42,48,54,60,66,72,78,84,90,96,102,108$.  The lattices have $4L^2/3$ sites.}
\label{fig:r2rho_anrecSU10}
\end{figure}

With adequate data in hand for a wide range of system sizes, we can attempt a careful and sophisticated collapse of the data.  We begin with the Binder ratio so as to extract $g_c$ and $\nu$ so that we can fix these values in dealing with other quantities of interest.  We include sub-leading corrections so as to fit to the following form:
\begin{equation}
\label{eqn:r2sub}
\mathbb{Y}_{R_2}(z)-(a+bz)L^{-\omega},
\end{equation}
where $\mathbb{Y}_{R_2}$ is just an analytic function of $z$, and $a$, $b$, and $\omega$ are fit parameters.  In practice, the data curves are very smooth since we have zoomed in considerably on the critical region and so we use a fifth order polynomial for $\mathbb{Y}_{R_2}$.  By minimizing the sum of the squares of the standard-error-weighted residuals between this polynomial and the Binder ratio data, shifted by the sub-leading corrections, the ideal values of $g_c$, $\nu$, $a$, $b$, and $\omega$ are chosen (note that this is a standard $\chi^2$ regression).  The parameter landscape has many shallow minima with the value of $\omega$ varying significantly but always of order unity.  We therefore fix $\omega$ at three different values: $\omega=0.5,1.0,2.0$.  Each value gives a different optimal $g_c,\nu$ pair.  Later, as each of these pairs, along with the corresponding value of $\omega$, are used to collapse the susceptibility data, we can use the variations in the optimal value of $\etaN$ to estimate its systematic error.  We can also collapse the spin stiffness, (specifically $\beta\rho_s$) to a similar scaling form with sub-leading corrections:
\begin{equation}
\label{eqn:rhosub}
\mathbb{Y}_{\rho}(z)-(a+bz)L^{-\omega}.
\end{equation}
Here, we fix the triplet $(g_c,\nu,\omega)$ using the results from the analysis of the Binder data and merely choose the optimal values of $a$ and $b$.  Figs.~\ref{fig:r2rho_hcSU5}~-~\ref{fig:r2rho_anrecSU10} show collapses of the Binder ratio and spin stiffness data for $N=5,7,10$ on the honeycomb and rectangular lattices.

Next, we come to the magnetic susceptibility data.  Here, we attempt to fit to the scaling form
\begin{equation}
\label{eqn:chiNsub}
L^{-1-\etaN}\left[\mathbb{Y}_{\text{N}}(z)-(a+bz)L^{-\omega}\right].
\end{equation}
We hold $g_c$, $\nu$, and $\omega$ fixed in the triplets found earlier and vary $a$, $b$, and $\etaN$ to find the optimal values.  We obtain a different value of $\etaN$ for each of the three triplets corresponding to $\omega=0.5,1.0,2.0$.  This yields an average and an upper and lower bound.  While we use this as an estimate of the systematic error, which appears as error bars in Fig.~5 of the main text, the true error is likely larger.

Now we consider the VBS susceptibility data.  Here, the stochastic error is greater than any potential corrections from sub-leading terms, so we neglect them in this case.  This almost certainly leads to systematic errors that are difficult to estimate with the available data.  What we can do reliably, however, is consider the difference in estimates of $\etaN$ when we turn off the sub-leading corrections.  This gives us an approximation of how much the anomalous scaling dimension can vary, percentage-wise, when we do not account for sub-leading corrections.  The error bars shown in the main text for $\etaV$ are the product of this approximation. The scaling form is
\begin{equation}
\label{eqn:chiVsub}
L^{-1-\etaV}\mathbb{Y}_{\text{V}}(z)
\end{equation}
and so we simply optimize for the parameter $\etaV$ with each of the $(g_c,\nu,\omega)$ triplets (even though there is no $\omega$ in the scaling form, there are still three separate pairs of $g_c$ and $\nu$).  Again, this gives an average, but in this case we do not use the variation in the three estimates to compute upper and lower bounds on $\etaV$; instead, we use the process described above for estimating the error bars in this quantity. It is notable that the estimates on the rectangular lattice are consistently lower than those on the honeycomb lattice, but it is clear from the large error bars that this difference could easily be accounted for by the inclusion of sub-leading corrections with less noisy data sets.

Finally, a brief mention of the situation regarding the exponent $\nu$.  This parameter is fitted during the collapse of the Binder ratio data.  While the fit values for SU(5) are clearly smaller than those for SU(7) and SU(10), the fit values for SU(7) and SU(10) do not differ greatly and, in some cases, are larger for SU(7) than SU(10).  This would seem to contradict the result from the field theory:
\begin{equation}
\label{eqn:nu}
\nu=1-\frac{48}{\pi^2 N}+\ldots.
\end{equation}
We can attempt to explain this discrepancy by considering that near the critical point, the quality of the data collapse is not strongly dependent on the scaling of $L$ in the rescaled variable $z$.  Hence, it is difficult to resolve clearly the value of $\nu$ in this regime.  We do see, however, that when we attempt to collapse Binder ratio data spanning a much wider range of values of $z$, such as in Fig.~\ref{fig:R2su5su10}, a monotonic progression of fit values for $\nu$ is indeed observed.  This suggests that perhaps a useful alternative approach to estimating $\nu$ as a first step is to use the wider view data.  Such an approach was not pursued here as this would likely result in a poorer estimate of $g_c$ and also because the anticipated impact on the estimates of $\etaN$ and $\etaV$ was small.

\end{document}